\documentclass[fleqn,usenatbib]{mnras}
\usepackage{newtxtext,newtxmath}
\usepackage[T1]{fontenc}
\DeclareRobustCommand{\VAN}[3]{#2}
\let\VANthebibliography\thebibliography
\def\thebibliography{\DeclareRobustCommand{\VAN}[3]{##3}\VANthebibliography}

\usepackage{graphicx}	
\usepackage{amsmath}	
\usepackage{multirow}
\usepackage{comment}
\usepackage{textcomp}

\title{Dust Substructures and Line Perturbations driven by a Forming Planet in J16120}

\author[A. Sierra et al.]{
Anibal Sierra$^{1,2}$\thanks{E-mail: asierra@astro.unam.mx}, 
Andrés F. Izquierdo$^{3,4}$,
Paola Pinilla$^{2}$,
Kan Chen$^{5}$,
Myriam Benisty$^{6}$,
Laura Pérez$^{7}$,
\newauthor
Teresa Paneque-Carreño$^{8}$,
Carolina Agurto-Gangas$^{9}$,
Jaehan Bae$^{3}$,
John Carpenter$^{10}$,
Luca Cieza$^{11,12}$,
\newauthor
Dingshan Deng$^{13}$,
Stefano Facchini$^{14}$,
Camilo González-Ruilova$^{12,15,16}$,
Nicolás Kurtovic$^{6,17}$,
\newauthor
Aleksandra Kuznetsova$^{18,19}$,
Carlo F. Manara$^{20}$,
James Miley$^{10,12,21}$,
Álvaro Ribas$^{22,23}$,
\newauthor
Giovanni Rosotti$^{14}$,
Miguel Vioque$^{20}$
\\
$^{1}$Universidad Nacional Autónoma de México. Instituto de Astronomía. A.P. 70-264, 04510. Ciudad de México, México\\
$^{2}$Mullard Space Science Laboratory, University College London, Holmbury St Mary, Dorking, Surrey RH5 6NT, UK\\
$^{3}$ Department of Astronomy, University of Florida, Gainesville, FL 32611, USA \\
$^{4}$ NASA Hubble Fellowship Program Sagan Fellow\\
$^{5}$ Kavli Institute for Astronomy and Astrophysics, Peking University, Beijing 100871, People’s Republic of China\\
$^{6}$Max-Planck Institute for Astronomy (MPIA), Königstuhl 17, 69117 Heidelberg, Germany \\
$^{7}$Departamento de Astronomía, Universidad de Chile, Camino El Observatorio 1515, Las Condes, Santiago, Chile\\
$^{8}$Department of Astronomy, University of Michigan, Ann Arbor, MI 48109, USA\\
$^{9}$Departamento de Física, Universidad Técnica Federico Santa María, Vicu\~{n}a Mackenna 3939, San Joaquín, Santiago de Chile, Chile \\
$^{10}$Joint ALMA Observatory, Alonso de Córdova 3107, Vitacura, Santiago 763-0355, Chile\\
$^{11}$Instituto de Estudios Astrofísicos, Universidad Diego Portales, Av. Ejercito 441, Santiago, Chile\\
$^{12}$Millennium Nucleus on Young Exoplanets and their Moons (YEMS), Chile\\
$^{13}$Lunar and Planetary Laboratory, the University of Arizona, Tucson, AZ 85721, USA\\
$^{14}$Dipartimento di Fisica, Università degli Studi di Milano, Via Celoria 16, Milano, Italy\\
$^{15}$Departamento de Física, Universidad de Santiago de Chile, Av. Victor Jara 3659, Santiago, Chile\\
$^{16}$Center for Interdisciplinary Research in Astrophysics and Space Science (CIRAS), Universidad de Santiago, Chile\\
$^{17}$Max Planck Institute for Extraterrestrial Physics, Giessenbachstrasse 1, D-85748 Garching, Germany\\
$^{18}$Center for Computational Astrophysics, Flatiron Institute, 162 Fifth Ave., New York, New York, 10025, USA\\
$^{19}$Department of Physics, University of Connecticut, 196A Auditorium Road, Unit 3046, Storrs, CT 06269, USA\\
$^{20}$European Southern Observatory, Karl-Schwarzschild-Strasse 2, D-85748 Garching bei München, Germany\\
$^{21}$European Southern Observatory, Alonso de Córdova 3107, Vitacura, Santiago, Chile\\
$^{22}$Astronomy Unit, Department of Physics and Astronomy, Queen Mary University of London, Mile End Road, London E1 4NS, UK\\
$^{23}$Institute of Astronomy, University of Cambridge, Madingley Road, Cambridge CB3 0HA, UK
}

\date{Accepted 2026 September 3. Received 2026 July 31; in original form 2026 June 9}

\pubyear{\the\year{}}

\begin{document}
\label{firstpage}
\pagerange{\pageref{firstpage}--\pageref{lastpage}}
\maketitle

\begin{abstract}
Hints of planet formation have been independently reported within the gap of the disc around 2MASS J16120668–301027 from millimetre continuum, infrared, and H$\alpha$ observations. In this work, we present new evidence for ongoing planet formation based on Atacama Large Millimeter/submillimeter Array (ALMA) Band 7 observations, detecting 0.87 mm dust continuum emission together with $^{12}$CO\,(J=3–2) and $^{13}$CO\,(J=3–2) line emission. Visibility modelling of the continuum data reveals an inner disc and two dust rings peaking at 23 and 75 au. The continuum morphology is better reproduced by an eccentric disc model ($e\sim0.1$) than by an axisymmetric disc. We further investigate the gas kinematics through modelling of the $^{12}$CO channel maps. The residual line-width map shows a localised increase in velocity dispersion at the position of a previously reported circumplanetary disc candidate (deprojected radius $\sim$32 au, position angle $\sim$170$^\circ$) and along its orbit. This signal is spatially coincident with kink-like features and a transition from sub-Keplerian to super-Keplerian velocities. In addition, the velocity residual map exhibits an arc-like structure extending outward from the planet candidate, while the gas kinematics—despite substantial uncertainties—is consistent with inflow towards the candidate's orbital radius. The observed increase in velocity dispersion agrees with predictions from planet–disc interaction simulations, which produce enhanced turbulence both at the planet location and along its orbital path. Taken together, the continuum morphology and gas kinematic signatures provide compelling new evidence for ongoing planet formation within the disc gap.
\end{abstract}

\begin{keywords}
Techniques: Interferometric -- Protoplanetary discs
\end{keywords}



\section{Introduction} \label{sec:introduction}

Protoplanetary discs have been the subject of extensive observational studies for several decades across a wide range of star-forming regions, providing key insights into the physical and chemical conditions under which planets form \citep[e.g.,][]{Beckwith_1990, Williams_2011, Dent_2013, Ansdell_2016, Pascucci_2016, Andrews_2018, Cieza_2019, Oberg_2021, Benisty_2023, Teague_2025, Zhang_2025}. These observations, spanning wavelengths from the infrared to the millimetre, have revealed a rich diversity of disc structures, including rings, gaps, spirals, and asymmetries \citep[e.g.,][]{Andrews_2018, Long_2018, Andrews_2020, Law_2020_rad, Bae_2023, Ginski_2024, Garufi_2024, Valegrard_2024, Guerra-Alvarado_2025}, many of which are thought to be linked to ongoing planet formation. In parallel, theoretical frameworks describing the formation and early evolution of planets have progressed from the foundational core accretion and gravitational instability models to more recent developments incorporating disc evolution, dust growth, and planet–disc interactions \citep[e.g.,][]{Goldreich_1973, Lissauer_1993, Pollack_1996, Liu_2020, Manara_2023, Orcajo_2025}.

Despite this progress, directly detecting forming planets within their natal discs remains observationally challenging. Young planets are expected to be deeply embedded and often obscured by optically thick dust and gas \citep{Sanchis_2020}, while their observational signatures can be subtle and easily confused with disc substructures produced by other mechanisms \citep{Currie_2015, Montesinos_2018, Szulagyi_2018, Szulagyi_2019, Szulagyi_2021}. As a result, robust confirmation typically requires the combination of multiple independent tracers, probing both dust and gas, and often spanning several wavelengths and observing techniques. Even though hundreds of discs have substructures, only a handful of proto-planets have been confirmed to date, specifically in the discs around PDS~70 \citep{Keppler_2018, Haffert_2019, Benisty_2021} and WISPIT~2 \citep{vanCapelleveen_2025, Close_2025}.
These rare systems provide crucial benchmarks for testing planet formation theories and motivate continued high-resolution, multi-tracer observations of discs with potential proto-planets.

Other systems in which tentative direct evidence for embedded planets has been suggested from infrared observations include AB~Aurigae \citep{Currie_2022}, and HD~169142 \citep{Hammond_2023}, where point sources have been identified within dust gaps potentially carved by the planets themselves. In contrast, systems showing indirect evidence, such as HD~163296 \citep{Teague_2021, Izquierdo_2022, Izquierdo_2026}, Elias~24 \citep{Pinte_2023}, HD~100546 \citep{Casassus_2019}, AS~209 \citep{Bae_2022, Izquierdo_2023}, exhibit either gas kinematic perturbations, Doppler flips, or chemical tracers sensitive to localised planet-induced heating. Additionally, the presence of a young massive planet has recently been proposed using Gaia astrometry combined with ALMA observations in the disc around MP Mus \citep{Ribas_2025}, demonstrating that giant planets can also be identified through proper-motion anomalies.


A recently proposed planet-host candidate studied across multiple wavelengths is the disc around 2MASS J16120668–301027 (hereafter J16120). ALMA Band 6 ($\lambda = 1.3$\,mm) dust continuum observations suggested the presence of a dusty circumplanetary disc (CPD) within a gap, at a radial separation of 32\,au from the central star \citep{Sierra_2024}; although higher angular resolution and sensitivity observations obtained one year later did not confirm this detection \citep{Li_2025}. In contrast, SPHERE infrared data \citep{Ginski_2025} revealed point sources in the H (1.6 $\mu$m) and K (2.2 $\mu$m) bands around the same location, and H$\alpha$ observations detected localised excess emission with a signal-to-noise ratio $\gtrsim 5$ \citep{Li_2025}. The latter may trace accretion onto a forming planet embedded within the gap, analogous to the H$\alpha$ features observed around planets PDS 70 b and c \citep{Haffert_2019, Thanathibodee_2019}.

Although the exact location of the planet candidate differs between tracers, \cite{Li_2025} demonstrated that the offset between the H$\alpha$ emission and the infrared point source in the K band can be explained by orbital motion over the two-year interval between the observations.
Together, these findings provide compelling evidence for the presence of a forming planet within the gap.
The mass of the planet candidate is estimated to lie between 0.1 and 5 M$_\mathrm{J}$, depending on the adopted models \citep{Ginski_2025}, which would be capable of carving the gap observed at millimetre and infrared wavelengths.

J16120 is an M0-M0.5 star with an age between 5 and 10 Myr, a luminosity of $0.25\,L_{\odot}$ \citep{Fang_2023, Ginski_2025}, an effective temperature of 3810 K, and an accretion rate of $\log \dot{M}_{\rm acc} = -9.21 \pm 0.5$ $M_{\odot}$ yr$^{-1}$ \citep{Ginski_2025}. Its X-Shooter spectrum is consistent with a $0.60 \pm 0.05$ M$_{\odot}$ star \citep{Ginski_2025}, while its dynamical mass estimated from the CO kinematics is $\sim 0.7\,M_{\odot}$ \citep{Sierra_2024}. All these properties are similar to the stellar properties of PDS~70 \citep{Muller_2018}. 

J16120 is located in the Upper Scorpius star-forming region at a distance of 132.1 pc \citep{Gaia_2023}, and its disc has previously been studied as part of the Upper Scorpius survey by \cite{Carpenter_2025} and the AGE-PRO ALMA Large Program \citep{Zhang_2025}. In the latter, the disc was detected in several molecular lines \citep{Agurto_2025}, including $^{12}$CO (J=2--1), $^{13}$CO (J=2--1), C$^{18}$O (J=2--1), N$_2$H+ (J=3--2), H$_2$CO (J=4$_{(0,4)}$--3$_{(0,3)}$), and C$^{34}$S (J=6--5), while it was not detected in DCN (J=4--3) and DCO$^{+}$ (J=4--3). From N$_2$H$^+$ and CO isotopologues, the gas disc mass was estimated to be $3.24^{+1.33}_{-0.72} \times 10^{-3}\,M_{\odot}$ \citep{Trapman_2025}, consistent with the external-photoevaporative viscous evolution models of \cite{Anania_2025}. The comparison between the dust and gas extents of J16120 indicates that radial drift has been halted or slowed by some mechanism \citep{Agurto_2025}, possibly linked to the ring-like structure, where dust grains may be trapped. In addition to the inner planet candidate, \cite{Sierra_2024} also reported a kink observed in several CO channel maps beyond the dust continuum emission. They speculated that the origin of this feature could be a second planet candidate in the outer disc.

In this paper, we present new high-sensitivity, high-angular-resolution, and high-spectral-resolution Band 7 ($\lambda = 0.87$,mm) observations of J16120 in dust continuum emission, $^{12}$CO ($J = 3$–2), and $^{13}$CO ($J = 3$–2), revealing new evidence of planet formation associated with the inner planet candidate in the gas emission. These observations reinforce earlier reports of a forming planet in the gap of J16120 and constitute a significant step forward in the detection of planetary signatures in protoplanetary discs.

\section{Observations}\label{sec:observations}
The disc around J16120 was observed by ALMA in Band 7 ($\lambda = 0.87$\,mm) from January to July 2024 (Project code: 2023.1.01100.S, PI: Anibal Sierra) in configurations C43-6 ($\sim 0.10^{\prime\prime}$) and C43-3 ($\sim 0.47^{\prime\prime}$) at a representative frequency of 	345.80 GHz. The former configuration consists of four long baseline (LB) executions, and the latter of three short baseline (SB) executions.
The spectral setup was configured with four spectral windows, with frequencies between 330.35 - 332.23 GHz, 330.55 - 330.61 GHz, 344.05 - 345.93 GHz, 345.76 - 345.82 GHz, and spectral resolution of 30.52 kHz and 976.56 kHz. This setup detects and spectrally resolves $^{12}$CO (J=3--2), $^{13}$CO (J=3--2), and dust continuum emission.

The continuum data and continuum-subtracted line emission were obtained following the imaging strategies for the discs in Upper Scorpius of AGE-PRO \citep{Zhang_2025, Agurto_2025}. The continuum-only datasets were obtained by identifying and flagging the molecular lines in each spectral window. 
All executions were aligned by fitting a 2D Gaussian to a low-angular-resolution image of the dust continuum in each execution, allowing us to identify the disc centre. We then re-centred all images to a common phase centre, chosen as J2000 16h12m06.664505s -30d10m27.61789s, to match the disc centre reported in Band 6 ($\lambda = 1.3$\,mm) in \cite{Sierra_2024}.

Before self-calibrating the data, we examined the dust continuum visibilities from each execution to identify any flux offsets between datasets. We found a flux difference of $\lesssim 10$\% between the short-baseline (SB) and long-baseline (LB) executions. The flux calibrators were observed closer in time to the LB executions\footnote{\url{https://almascience.eso.org/sc/}} ($\sim$2 days of difference for executions uid://a002/X1198c0c/X2af, uid://a002/X1199f9e/X528, and uid://a002/X1199f9e/X7f3a) compared to the SB executions ($\sim$10 days of difference). Therefore, we adopt the flux of the concatenated LB executions as reference and scale the fluxes of all executions accordingly using the task \texttt{gencal}. We double-check that de-coherence is not an important issue before applying the flux-rescaling.

Self-calibration was performed using CASA \citep{McMullin_2007} version 6.4.3.27. We first self-calibrated the SB data only. After two rounds of phase self-calibration where all the SB spectral windows and scans were combined using the task \texttt{gaingal} using \texttt{coombine='spws, scans'}, time interval solutions \texttt{solint='inf', '360s'}, and calibrating the data only using the task \texttt{applymode='calonly'}, the signal-to-noise (SNR) improved by a factor of 143\% and 8\% with respect to the previous iteration, respectively.
Then, we concatenate the LB executions and proceed to self-cal the whole dataset. In this process, the antenna reference is taken as that with the best SNR, as reported by the weblog. After one round of phase self-calibration where spectral windows and scans were combined and time interval solution of \texttt{solint='inf'} was applied, the SNR increased by 5.6\%. Additional tests using shorter solution intervals and amplitude self-calibration did not result in significant SNR improvements, so these were not applied to the final products. The final dust continuum image has an angular resolution of $0.125^{\prime\prime} \times 0.095^{\prime\prime}$, and a SNR of 69 when imaged with robust\,=\,0.5. Compared to the dust continuum image in Band 6 \citep{Sierra_2024}, both beam area and SNR of the Band 7 image in this work are smaller by a factor of 2.

The molecular line data were obtained by applying the same astrometric aligns, flux offset scales, and self-calibration tables to the original datasets without channel averaging, and subtracting the dust continuum using the task \texttt{uvcontsub}. The final datasets are resample onto a regular velocity grid using the task \texttt{cvel2}, with a spectral resolution of 25 m s$^{-1}$.

The dust continuum, $^{12}$CO, and $^{13}$CO datasets were imaged using the \texttt{tclean} task in CASA, employing the multi-scale deconvolution algorithm with scales corresponding to point sources, one beam size, and two beam sizes.
For the molecular lines, we generated a grid of image cubes with varying spectral resolutions (25, 50, and 100 m s$^{-1}$) and Briggs robust parameters ($r = 0.0, 0.5, 1.0, 1.5, 2.0$), aiming to identify the optimal combination where small-scale spatial structures can be observed with good SNR in individual channel maps.
The fiducial products were generated with a spectral resolution of 25\,m\,s$^{-1}$ and a Briggs robust parameter of 1.5 for $^{12}$CO, and 50\,m\,s$^{-1}$ with a robust parameter of 2.0 for $^{13}$CO. For the dust continuum, we adopted a robust parameter of 0.5. The JvM correction \cite{Jorsater_1995, Czekala_2021} was not applied to any of the final products; however, we cleaned down to a 1$\sigma$ threshold, with very little signal not included in the clean model and therefore affected by the non-Gaussianity of the beam \citep{Zhang_2025}. Table \ref{tab:Obs} shows a summary of the image parameter of the fiducial products for the Band 7 observations.

Additionally, we use the self-calibrated Band~6 observations of J16120 presented in \cite{Li_2025}, in which data from programs 2021.1.00128.L \citep[PI: Ke Zhang,][]{Zhang_2025} and 2022.1.00646.S (PI: Feng Long) were concatenated to produce the final dust continuum products. We refer the reader to Appendix~A of \cite{Li_2025} for details of the Band~6 self-calibration procedure.

\begin{table}
    \centering
    \caption{Band 7 image parameters.}    
    \begin{tabular}{ccccc}
    \hline 
    Data  &  Robust & Beam & rms$^{*}$ & VRes \\
    &   & [mas $\times$ mas; deg] & [mJy beam$^{-1}]$ & [m s$^{-1}$]\\
    \hline
    Continuum & 0.5 & 125 $\times$ 95; -75.2 & 0.021& -\\ 
    $^{12}$CO (J=3--2) & 1.5 & 181 $\times$ 145; -78.8 & 4.9 & 25\\
    $^{13}$CO (J=3--2) & 2.0 & 183 $\times$ 143; -75.6 & 4.0 & 50\\
    \hline
    \end{tabular}
    $^{*}$For line maps, this is the rms per channel. 
    \label{tab:Obs}
\end{table}

\begin{table}
    \centering
    \caption{Disc Sizes and Integrated Fluxes of J16120.}    
    \begin{tabular}{ccccc}
    \hline 
    Data  &  $R_{90}$ & $R_{68}$ & Flux \\
            &  [au] & [au] & [mJy km/s]$^{*}$ \\    
    \hline
    Continuum ($\lambda = 0.87$ mm) & 95 $\pm$ 14          & 81 $\pm$ 14 & 68 $\pm$ 3\\
    $^{13}$CO (J=3--2) & 140 $\pm$ 20 & 113 $\pm$ 20 & 2600 $\pm$ 300\\    
    $^{12}$CO (J=3--2) & 180 $\pm$ 20 & 123 $\pm$ 20 & 10600 $\pm$ 500\\    
        \hline
    \end{tabular}
    
    $^{*}$For continuum, units are mJy.
    \label{tab:Properties}
\end{table}

\section{Results}\label{sec:results}
\subsection{Dust continuum and molecular moment maps}

\begin{figure*}
    \centering
    \includegraphics[width=\linewidth]{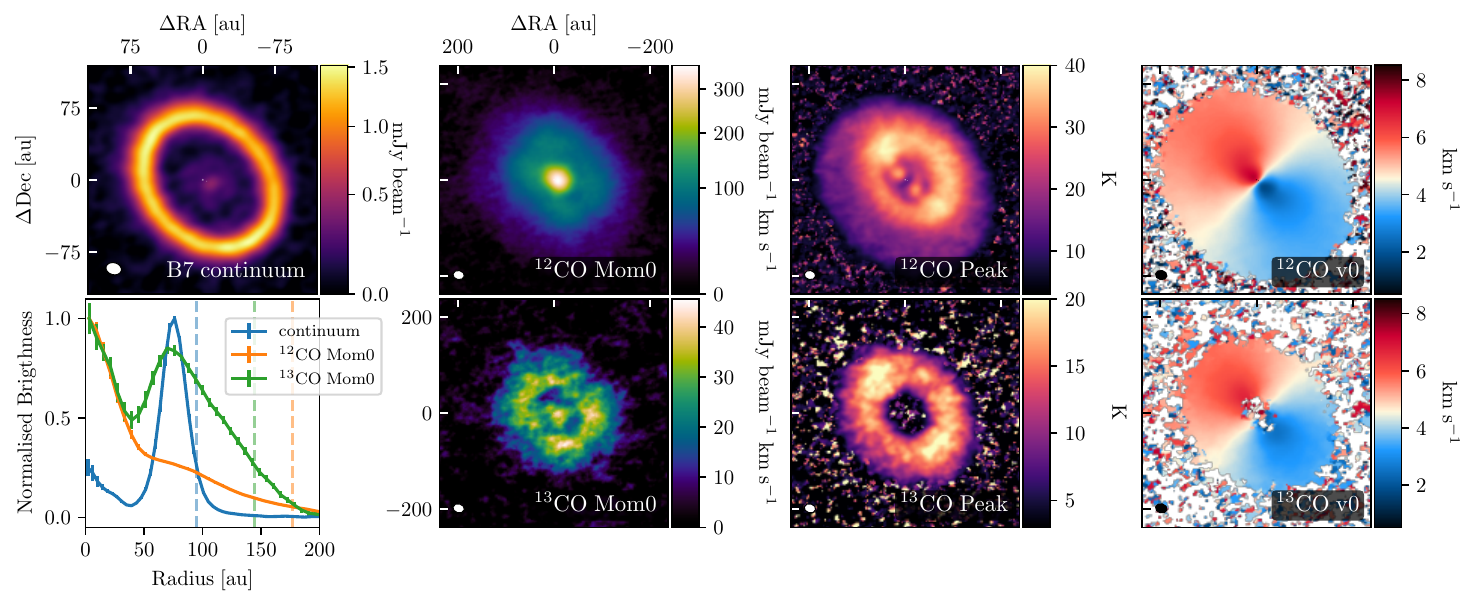}
    \caption{ALMA Band 7 ($\lambda = 0.87$\,mm) observations of the disc around J16120. First row: Dust continuum map (left), Moment 0 (middle-left), peak intensity (middle-right), and Gaussian velocity map (right) map of the $^{12}$CO emission.
    Second row: Normalised brightness profiles computed from the integrated Moment 0 maps (left), Moment 0 (middle-left), peak intensity (middle-right), and Gaussian velocity map (right) map of the $^{13}$CO emission. The vertical dashed lines in the left panel show the $R_{90}$ radius of each tracer.}
    \label{fig:Observations}
\end{figure*}

The map of the dust continuum emission, and moment maps \citep[computed using \textsc{Discminer, }][]{Izquierdo_2021} from $^{12}$CO, and $^{13}$CO are presented in Figure \ref{fig:Observations}. The moments were computed by clipping the data to 1$\sigma$ for Moment 0 (Integrated intensity map), and $3 \sigma$ for the peak intensity map, and the peak velocity.

The disc presents a ring like morphology and a bright inner disc in both dust and $^{13}$CO emission, as revealed by the moment 0 brightness radial profile. The peaks of the dust continuum and the $^{13}$CO emission coincide, likely indicating an accumulation of both dust and gas in this region. The observed morphology is consistent with previous millimetre and infrared observations \citep{Sierra_2024, Ginski_2025,  Li_2025}. The central emission in $^{12}$CO is more homogeneous (probably due to optical depth effects) but still presents evidence of a ring and a bright central component. The peak brightness reveals different disc morphologies in $^{12}$CO and $^{13}$CO. In the optically thick regime (as typically expected for $^{12}$CO), the peak intensity approximates the brightness of the emitting layer. However, in the optically thin case (more representative of $^{13}$CO), the peak intensity traces a combination of gas surface density and temperature (at lower vertical heights compared to $^{12}$CO).

The circumplanetary disc (CPD) candidate observed in Band 6 ($\lambda = 1.3$\,mm) by \cite{Sierra_2024} in the centre of the gap is not observed in the Band 7 ($\lambda = 0.87$\,mm) data. Both observations have a sensitivity of 0.021 mJy beam$^{-1}$ (or $\sim 20$ mK). This CPD candidate was also not recovered in the deeper observations in Band 6 by \cite{Li_2025}. In \cite{Sierra_2024}, the recovery fraction (the fraction of point sources that are recovered from injection test) of such CPD candidate in the disc gap is only $\sim 0.2$. Consequently, the evidence for dust continuum compact emission at the centre of the gap in J16120 is unlikely to be robust and may be associated to an imaging artifact or thermal noise. Estimations on the Band 7 detectability based on the expected spectral index are discussed in Section \ref{sec:NoB7Detection}.

The line-of-sight velocity exhibits the characteristic blue- and red-shifted pattern of a rotating disc and is globally consistent with Keplerian motion (Section \ref{sec:gas_kinematics}). The disc radii enclosing 90\% and 68\% of the total flux ($R_{90}$ and $R_{68}$, respectively), and fluxes of each tracer are reported in Table \ref{tab:Properties}. These values were derived using the emission curve of growth (COG), and omitting emission at radii greater than $4^{\prime\prime}$ from the disc centre to avoid contamination from a millimetre dust continuum background source (Appendix \ref{sec:background}). Although there is not gas emission associated to the background source, we also omit the emission beyond $4^{\prime\prime}$ for the gas tracers. This choice does not affect the measured radii, as it is at least a factor of 3 larger than the measured CO radii.

Both the dust continuum and $^{12}$CO radii are consistent (within the error bars and taking into account the beam difference) with the values reported in \cite{Agurto_2025} at ALMA Band 6. The $^{12}$CO and $^{13}$CO radii are approximately $1.9$ and $1.5$ times larger than the dust-continuum radii, respectively, implying that this source is not affected by efficient radial drift \citep{Facchini_2017, Trapman_2019, Toci_2021, Pinilla_2025}.
The morphology of the dust continuum and line emission are studied in Sections \ref{sec:Dust_continuum_analysis} and \ref{sec:Gas_emission} respectively.

\subsection{Dust continuum emission morphology}
\label{sec:Dust_continuum_analysis}

The dust continuum map of J16120 presents two clear bright spots along the semi-major axis of the disc. This may arise from disc azimuthal asymmetries or from an optically thin and geometrically thick disc \citep{Doi_2021}. Additionally, the disc is known to exhibit significant azimuthal asymmetries that appear to follow a $m=1$ spiral arm structure \citep{Sierra_2024}.
Therefore, to explore both scenarios, we study the dust continuum morphology in the visibility plane, assuming both an axisymmetric disc (Section \ref{sec:Frank}) and an eccentric disc (Section \ref{sec:Galario}). In the former, we use \textsc{frankenstein} \citep{Jennings_2020}, which assumes and axisymmetric disc, and in the latter we use \textsc{Galario} \citep{Tazzari_2018} to explore the free parameters that describe the eccentric disc morphology.
In both cases, the continuum visibilities were extracted from the self-calibrated measurement sets using the function \textit{export\_uvtable} in \cite{uvplot_tazzari}, but normalising the uv-distances with the wavelength of each spectral windows and channel. 

Before modelling the visibilities, we compute constraints to the disc geometry by minimizing the spread of the real part of the visibilities \citep{Isella_2019} in both Band 6 and 7. We obtained an inclination of $\rm inc$=$36.9^{+0.7}_{-0.1}$\,deg, and a position angle of PA=$45.0^{+0.7}_{-0.2}$\,deg in Band 7, and inc=$37.0 ^{+0.1}_{-0.2}$\,deg, PA=$45.1^{+0.2}_{-0.9}$\,deg in Band 6. These values are identical. From here on out, we use the inclination and position angle inferred from the optically thinner Band 6 as the geometric constraints. 

\subsubsection{Axisymmetric dust disc}
\label{sec:Frank}
\begin{figure*}
    \centering
    \includegraphics[width=\linewidth]{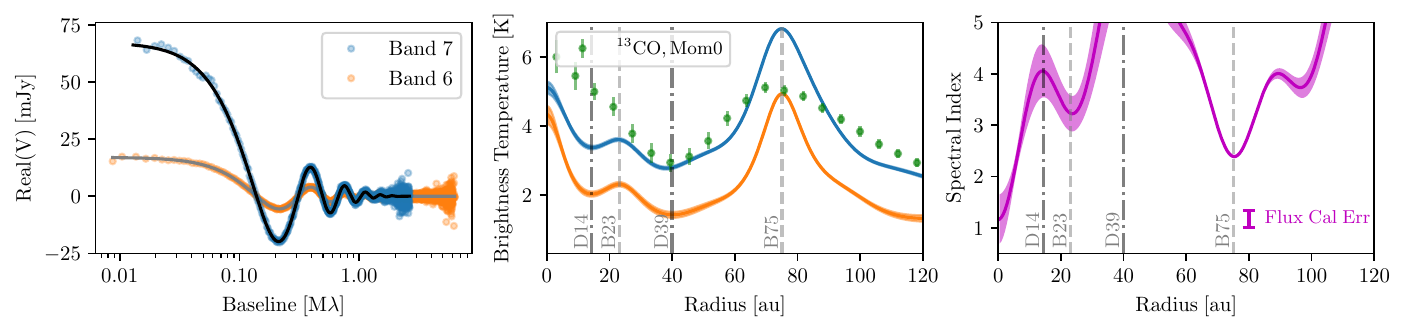}
    \caption{Left: Visibilities of the dust continuum emission at Band 7 ($\lambda = 0.87$\,mm, blue) and 6 ($\lambda = 0.87$\,mm, orange). The \textsc{frankenstein} models are shown as a solid line. Middle: Brightness radial profiles obtained from visibility fit, alongside the $^{13}$CO Moment 0 radial profile scaled for visual comparison on the same vertical axis. Right: Spectral index between Band 7 and 6. The error bar in the bottom right corner is the Flux calibrator error assuming 10\% of flux uncertainty in each band. The vertical axis is truncated at $\alpha = 5$, as the spectral index attains very high values in the vicinity of the deep dust continuum gap.
    The dashed/dashed-dotted vertical lines in the middle and right panel indicate the position of the rings/gaps.}
    \label{fig:Vis_Res_SPI}
\end{figure*}

The \textsc{Frankenstein} visibility modelling of the self-calibrated dust continuum visibilities of the disc around J16120 at ALMA Band 6 ($\lambda = 1.3$\,mm) and 7 ($\lambda = 0.87$\,mm) are shown in the left panel of Figure \ref{fig:Vis_Res_SPI}. 
In both cases, the disc centres were computed by minimizing the imaginary part of the visibilities, obtaining an offset in Right Ascension and Declination of ($\Delta$RA, $\Delta$Dec) = ($-27.1^{+2.4}_{-0.1}$  mas, $-22.3^{+0.4}_{-0.1}$ mas) for Band 7, and ($-11.6^{+0.8}_{-0.6}$ mas, $-1.6^{+0.9}_{-0.4}$ mas) for Band 6. Additionally, in both cases we fit the visibilities using the hyper-parameters $\alpha = 1.05$, and $w_{\rm smooth} = 0.1$. The former guarantees that low–signal-to-noise data are incorporated into the fit, whereas the latter represents a conservative choice that mitigates over-fitting of the visibility power spectrum. Tests with various hyper-parameter combinations revealed no significant differences.
The middle panel of Figure \ref{fig:Vis_Res_SPI} shows the brightness temperature radial profile obtained from the \textsc{Frankenstein} fit, without assuming the Rayleigh-Jeans approximation. Two bright rings at 23 and 75 au (B23, B75, respectively) are independently inferred in both wavelengths in addition to the inner bright disc, and two gaps at 14 and 39 au (D14, D39, respectively). The D39 dust gap also coincides with a gas gap partially observed in $^{12}$CO, and fully observed in $^{13}$CO. The B23 ring was already reported in the visibility modelling of \cite{Li_2025}. 

The right panel in Figure \ref{fig:Vis_Res_SPI} presents the radial spectral index profile between the two wavelengths, revealing local minima associated with the bright rings and local maxima located in the gaps. The spectral index in the innermost region ($<5$ au) is below 2, suggesting very optically thick emission and important scattering effects \citep{Zhu_2019, Sierra_2020} or free-free contamination \citep{Rota_2024}, although the latter is not expected to contribute significantly at ALMA Band 6 and 7 wavelengths. Outside this region, the spectral index is above 3, except around B75, where it is $\sim 2.3$. In the region around the D39 gap, where the disc emission in both Band 7 and Band 6 is on the order of the thermal noise, the spectral index reaches unphysical high values ($\sim7$).
All these values are subject to uncertainties in the flux calibration, which can vertically shift the spectral index radial profile by $\sim 0.5$.

We also use the multi-wavelength modelling described in \cite{Sierra_2024b} to study the distribution of solids (such as maximum grain size, temperature, and dust surface density, e.g., \citealt{Carrasco-Gonzalez_2019, Macias_2021, Sierra_2021, Viscardi_2025}), and we find that the posterior probabilities of both the maximum grain size and the dust surface density tend to peak values at the locations of the bright rings, suggesting the presence of dust traps in B23 and B75. The spectral index also shows local minima at the same locations, which is also indicative of dust growth in pressure bumps or dust traps \citep[e.g.,][]{Pinilla_2012b}. However, as discussed in \cite{Viscardi_2025}, using only two millimetre wavelengths is insufficient to constrain the dust properties due to significant degeneracies. At least one additional independent dust continuum observation at a different wavelength is therefore required to confirm these trends. Moreover, in Section~\ref{sec:gas_kinematics}, we suggest that the gas around B23 may be flowing from the outer disc towards this orbit, further supporting the dust trap scenario.

\begin{figure}
    \centering
    \includegraphics[width=\linewidth]{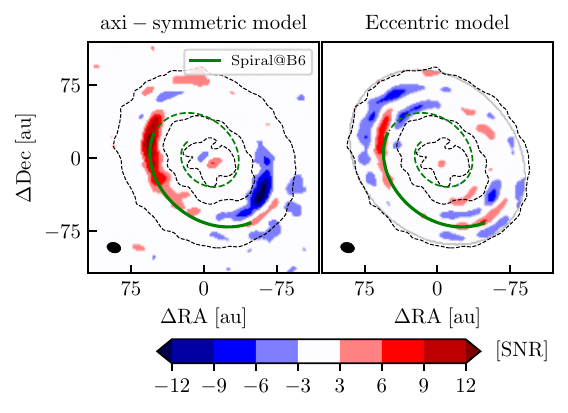}
    \caption{Band 7 ($\lambda = 0.87$\,mm) dust continuum residual map after subtracting the best-fit model from \textsc{Frankenstein} (left panel) and eccentric parametric model with \textsc{Galario} (right panel). The green line in both panels shows the spiral structure observed in the residual maps of the best fit model using Band 6 data \citep{Sierra_2024}. The iso-contours show the dust continuum ALMA Band 7 data at 7-$\sigma$ level.}
    \label{fig:Residual_maps}
\end{figure}

The left panel of Figure \ref{fig:Residual_maps} shows the map of the residual visibilities (Data - Frankestein fit). This map shows a positive residual structure in the Eastern region of the ring, and a negative residual in the West, similar to the residual map at Band 6 in \cite{Sierra_2024, Vioque_2025a, Li_2025}. In the former, a spiral function was fitted to the positive residuals. This possible spiral spatially coincides with the positive residuals observed in Band 7 (green lines), suggesting that this apparent spiral is not an artifact \citep[e.g., by errors in the disc geometry or offset, ][]{Andrews_2021}, but a real non-axisymmetric structure. In the next section we show that some of this non-axisymmetric structure can also be described by a non-eccentric disc.

\subsubsection{Eccentric dust disc}
\label{sec:Galario}

Motivated by the residuals obtained after subtracting an axisymmetric model, we explore in this section a more sophisticated non-axisymmetric disc morphology. In particular, we describe the disc morphology using an eccentric disc following the eccentric coordinates implemented in previous works \citep[e.g.,][]{Marino_2019, Booth_2021, Kurtovic_2022}, where the semi-major axis is given by:

\begin{equation}
    a(r,\phi) = r \left( \frac{1 -e \cos(\phi - \omega)}{1-e^2} \right),
\end{equation}
where $r, \phi$ are the radial and azimuthal coordinates measured relative to the disc centre and the disc minor axis in the North-West, respectively, $e$ is the disc eccentricity, and $\omega$ is the periastron. 
We use the Python library \textsc{emcee} introduced in \cite{Foreman_2013} to explore the posterior distributions of the parameters describing the disc model, and \textsc{Galario} to compute their visibilities. The pixel size and number of pixels are fixed to 10 mas and 1024, respectively. The inclination and position angle are fixed at 37.0 deg and 45.1 deg, respectively, to avoid degeneracies with the disc periastron.

The disc model consists of three Gaussians: $g_0$, $g_{23}$, $g_{75}$, which mimic the ring like structure inferred from the axisymmetric disc. The centre ($r_i$), amplitude ($A_i$), and width ($\sigma_i$) of each Gaussian are free parameters of the model, in addition to the disc offset ($\Delta$RA, $\Delta$Dec), the eccentricity ($e$), and periastron ($\omega$), resulting in a total of 13 free parameters. The explored parameter space for the Gaussian components are motivated by the results of the axisymmetric disc. For the eccentricity, we explore values between 0 and 1, and for the periastron between $-\pi$ and $\pi$. In all cases, we adopt uniform (linear) priors. The number of walkers is set to eight times the number of free parameters. Convergence is assessed by inspecting the walker chains and their autocorrelation times after 8000 steps. The best-fit parameter values (the one that minimises the chi-squared) are summarised in Appendix~\ref{app:best-fits}.

The residual map of the eccentric model is presented in the right panel of Figure \ref{fig:Residual_maps}. Unlike the residuals from the axisymmetric disc model, the eccentric model better reproduces the continuum asymmetries.
We also tested models that include additional rings in the fit, but neither the residual maps nor the chi-squared values show significant improvement. This simple eccentric three-ring model does a better job at reproducing the disc morphology (in particular, the azimuthal asymmetries) compared to the axisymmetric disc. However, some of the spiral-like residuals present in the axisymmetric model remain, leaving open the possibility that a spiral is present.

\subsection{Gas emission morphology}
\label{sec:Gas_emission}


Gas disc morphology is influenced by the presence of embedded planets \citep{Kley_2012}. The high angular resolution and sensitivity of the observations of the $^{12}$CO (J=3--2) and $^{13}$CO (J=3--2) emission around J16120 motivate an investigation into the origin of the observed gas substructures and their potential connection to embedded planet candidates.

\begin{figure*}
    \includegraphics[width=\linewidth]{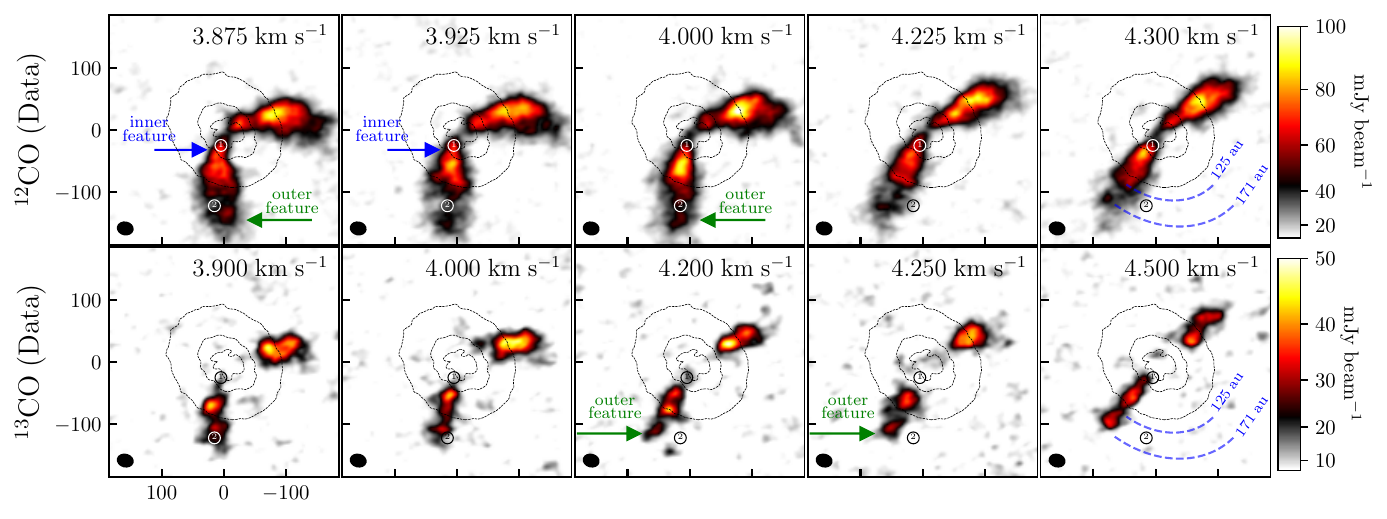}
    \caption{Continuum-subtracted selected $^{12}$CO (top panels) and $^{13}$CO (bottom panels) channel maps showing gas features in the inner disc (blue arrows) and in the outer disc (green arrows). The latter may arise from projection effects between the front and back sides of the disc, or a gas gap. The position of the inner planet candidate at $r = 32$ au is marked by \protect\textcircled{\textbf{1}}, while that of the outer planet candidate at $r = 148$ au is marked by \protect\textcircled{\textbf{2}}. The radial range between 125 and 171 au, where the outer feature is observed, is marked in the right panels. Colour bar in all panels starts at $1\sigma$. The iso-contours show the ALMA Band 7 dust continuum emission at 7-$\sigma$ level.}
    \label{fig:CO-channels}
\end{figure*}

Selected channel maps (within 3.875 - 4.500 km s$^{-1}$) are shown in Figure ~\ref{fig:CO-channels}. The $^{13}$CO emission shows low intensity within the cavity, consistent with the presence of a gas cavity as revealed by the moment 0 map (Figure \ref{fig:Observations}). The $^{12}$CO emission extends towards the inner disc, with some asymmetries, or kink-like features, at an orbital radius of $\sim 50$\,au. As these features lie in the vicinity of the inner planet candidate proposed in \cite{Sierra_2024} at 32\,au, they could be associated with its presence. After modelling the disc kinematics, we discuss the significance of these features in more detail in Section~\ref{sec:Discussion_Gas}.

On the other hand, we identify another intensity feature resembling a kink-like morphology in the outer disc, possibly associated with kinematic perturbations, toward the southern side of the disc in both molecular tracers. These structures are located between radii of $\sim125$ au and $\sim171$ au (Hereafter, we refer to the associated planet candidate as the 148 au planet candidate, corresponding to the mean radius of these structures), far beyond the outer edge of the dust continuum emission. In \cite{Sierra_2024}, these structures were reported to lie at a de-projected radius of $\sim\,112$au based on the velocity residual map. However, the current observations show that they extend significantly farther out.

In some channels, particularly in $^{13}$CO, the features appear partially detached from the global disc emission and resemble the intensity signatures expected from planet-disc interactions \citep[e.g.,][]{Izquierdo_2021} or from a gaseous circumplanetary disc \citep[e.g.,][]{Bae_2022}.  However, the location where these structures are observed varies significantly between channels, making these interpretations less plausible.
In Section~\ref{sec:Discussion_Gas}, we discuss how these features may be explained by projection effects arising from the front and back sides of the disc, although they may also trace a gas gap. We also assess the significance of these features with respect to the best-fit Keplerian model.

\subsubsection{Disc kinematics}
\label{sec:gas_kinematics}

The high sensitivity, spectral resolution, and angular resolution of the gas observations toward J16120 make this dataset exceptionally well suited for kinematic modelling. We use \textsc{discminer} \citep{Izquierdo_2021} to fit the intensities and rotation velocities encoded in the channel maps of the $^{12}$CO (J=3--2) and $^{13}$CO (J=3--2) emission from this target.

Following \citet{Izquierdo_2025}, our models adopt smooth parametric prescriptions to describe the radial variations of the line-profile properties of each tracer, including the peak intensity, line width, and line slope. The velocity shifts of the model lines are computed assuming Keplerian motion set by the stellar mass $M_\star$, with differential rotation as a function of height above the disc midplane. We also include the systemic velocity of the source, $\upsilon_{\rm sys}$, as a free parameter. The models further account for the projected geometrical structure of the channel maps, defined by the disc inclination $i$ and position angle PA (both treated as free parameters, with best-fit values consistent with those inferred from the dust continuum analysis; Section~\ref{sec:Dust_continuum_analysis}), as well as the emission surface height $z(r)$ in the front and back sides of the disc, independently parametrised as exponentially tapered radial profiles (see Eq. \ref{eq:surface}). 

We explore the parameter space using the MCMC ensemble sampler \textsc{emcee} \citep{Foreman_2013}, employing 230 walker chains, burn-in phases of 80,000 and 30,000 steps for the $^{12}$CO and $^{13}$CO cubes, respectively, and an additional 10\% of steps to sample the posterior distributions. To study the presence of perturbations in the disc, we compare the best-fit model channels with those of the data to extract residual maps that reveal deviations from Keplerian rotation, as well as gas substructures manifested as intensity and line-width fluctuations. Because the backside of the disc does not contribute significantly to the observed emission \citep[see discussion in ][or model images in Figure \ref{fig:CO-channels_Model}]{Izquierdo_2025}, we adopt Gaussian profiles to estimate centroid velocity, peak intensity, and line-width maps from both the data and the corresponding \textsc{discminer} model. The best fit parameters are summarised in Appendix \ref{app:best-fits}.

\begin{figure*}
    \centering
    \includegraphics[width=\linewidth]
    {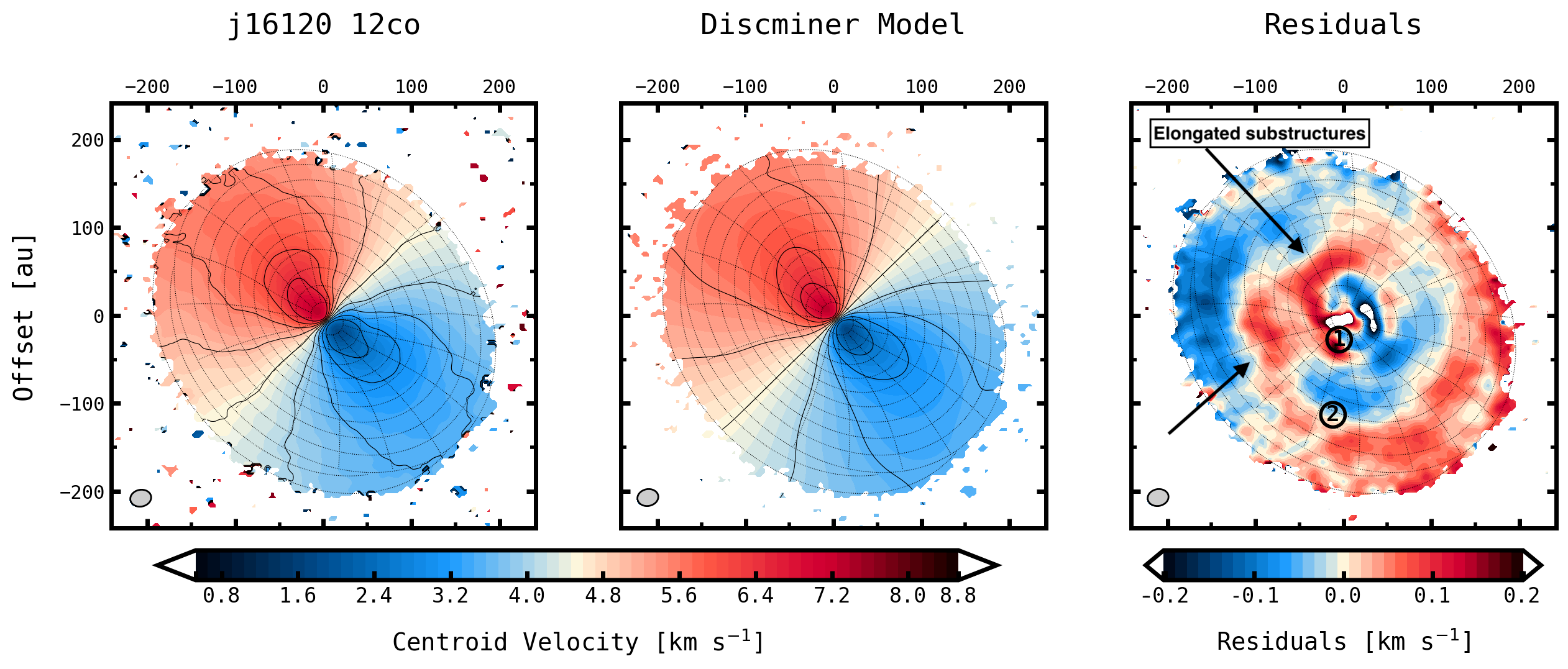}
    \caption{Kinematics of the $^{12}$CO emission around J16120. Left: Peak velocity along the line of sight. Middle: Discminer model. Right: Residual map. 
    The position of the inner planet candidate at $r = 32$ au is marked by \protect\textcircled{\textbf{1}}, while that of the outer planet candidate at $r = 148$ au is marked by \protect\textcircled{\textbf{2}}. The arrows indicate the location of elongated residual substructures.
    }
    \label{fig:CO-Discminer_Residuals}
\end{figure*}

Selected channel maps of the disc model and residuals are shown in Figure \ref{fig:CO-channels_Model}. The results for the velocity along the line of sight for $^{12}$CO are shown in Figure \ref{fig:CO-Discminer_Residuals}, and in Figure \ref{fig:13CO-Discminer_Residuals} for $^{13}$CO. In both tracers, the disc model successfully reproduces the overall rotation pattern of the system.

The channel residual maps (bottom panels of Figure \ref{fig:CO-channels_Model}) show that the intensity residuals around the outer kink-like feature are generally not significant. 
Although a few residual structures remain (e.g., $^{13}$CO at 4.500\,km\,s$^{-1}$, where the peak residual reaches 20.6\,mJy\,beam$^{-1}$, corresponding to an SNR of $\sim$5), in most channels, including those around the kink-like features, the residuals render these features consistent with the measured image noise.
Similar substructures have been reported in the channel maps of Elias~2-27 \citep{Paneque_2021}, where they were not interpreted as deviations from Keplerian rotation. Instead, that study attributed the emission to projection effects from the upper and lower disc surfaces, connected by material bridging the midplane between the two sides, although they could also have been associated with gravitational instability.

On the other hand, the $^{12}$CO velocity residuals (right panel of Figure \ref{fig:CO-Discminer_Residuals}) reveal two elongated substructures extending toward the inner and outer regions of the disc, closely overlapping with the locations of the planet candidates proposed by \cite{Sierra_2024}. No comparable structures are observed in the $^{13}$CO residuals; however, the lower sensitivity of this line makes it harder to assess the presence or absence of signatures associated with the proposed planet candidates.

\begin{figure*}
    \centering
    \includegraphics[width=0.3\linewidth]{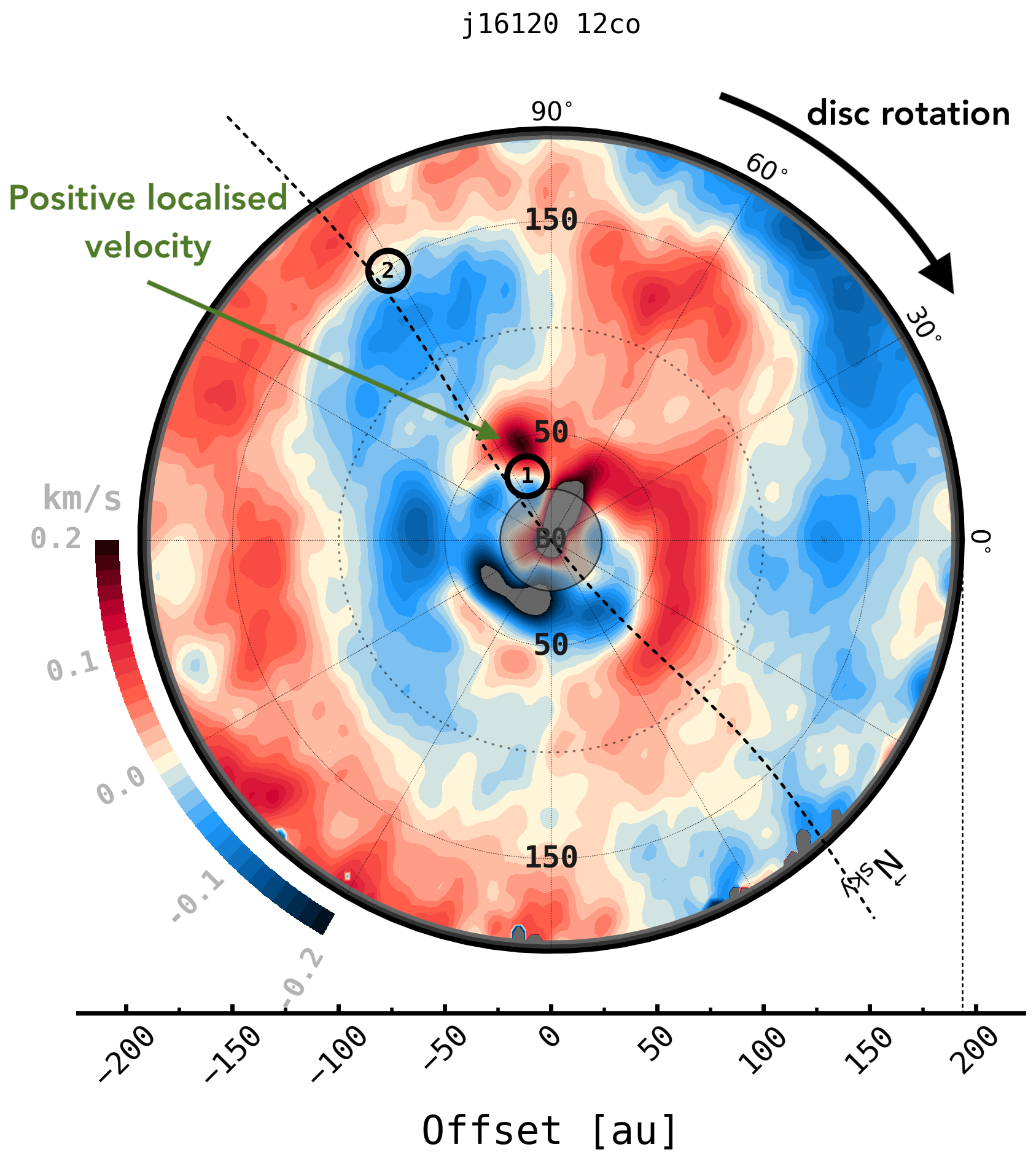}
    \includegraphics[width=0.3\linewidth]{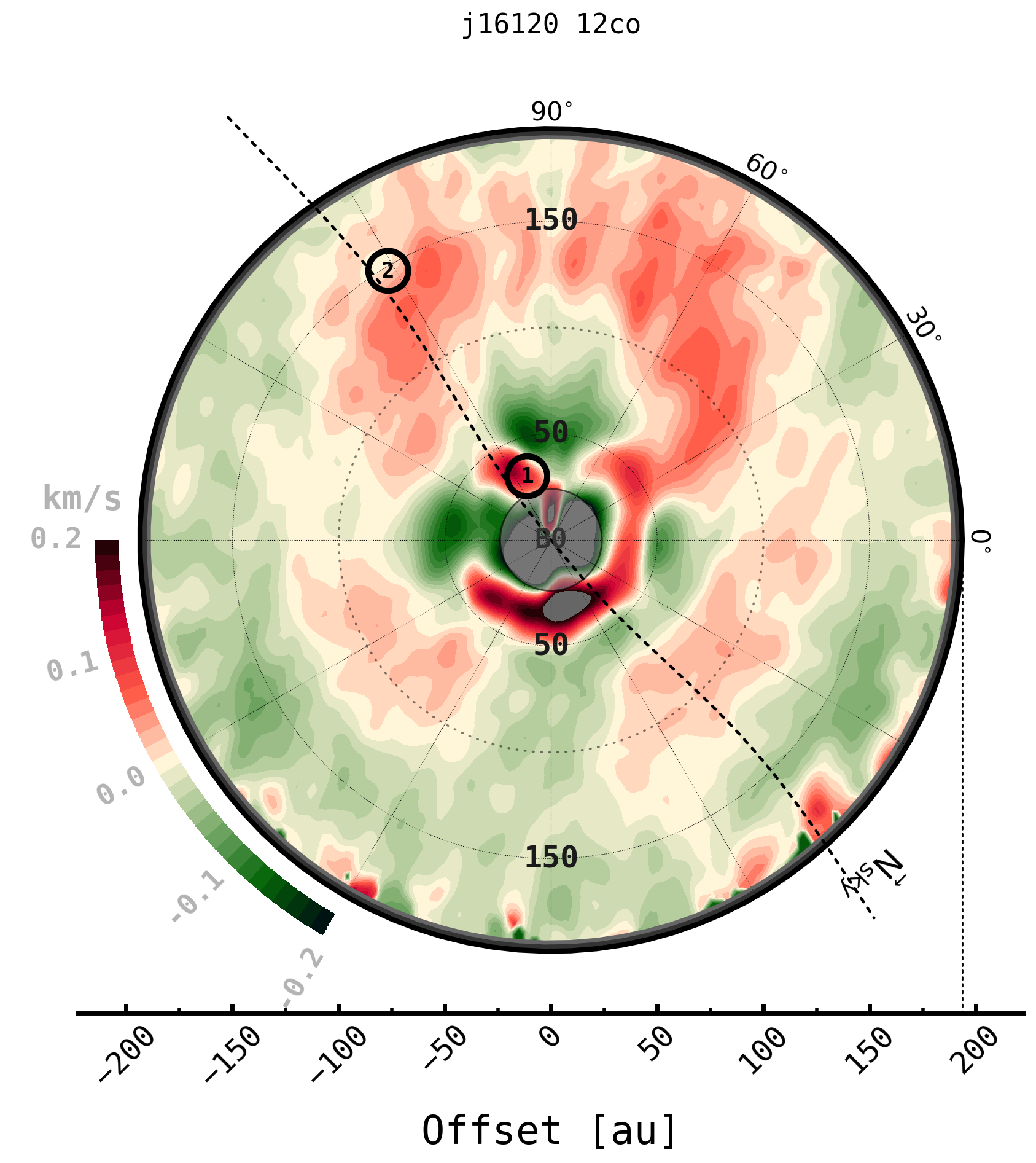}
    \includegraphics[width=0.3\linewidth]{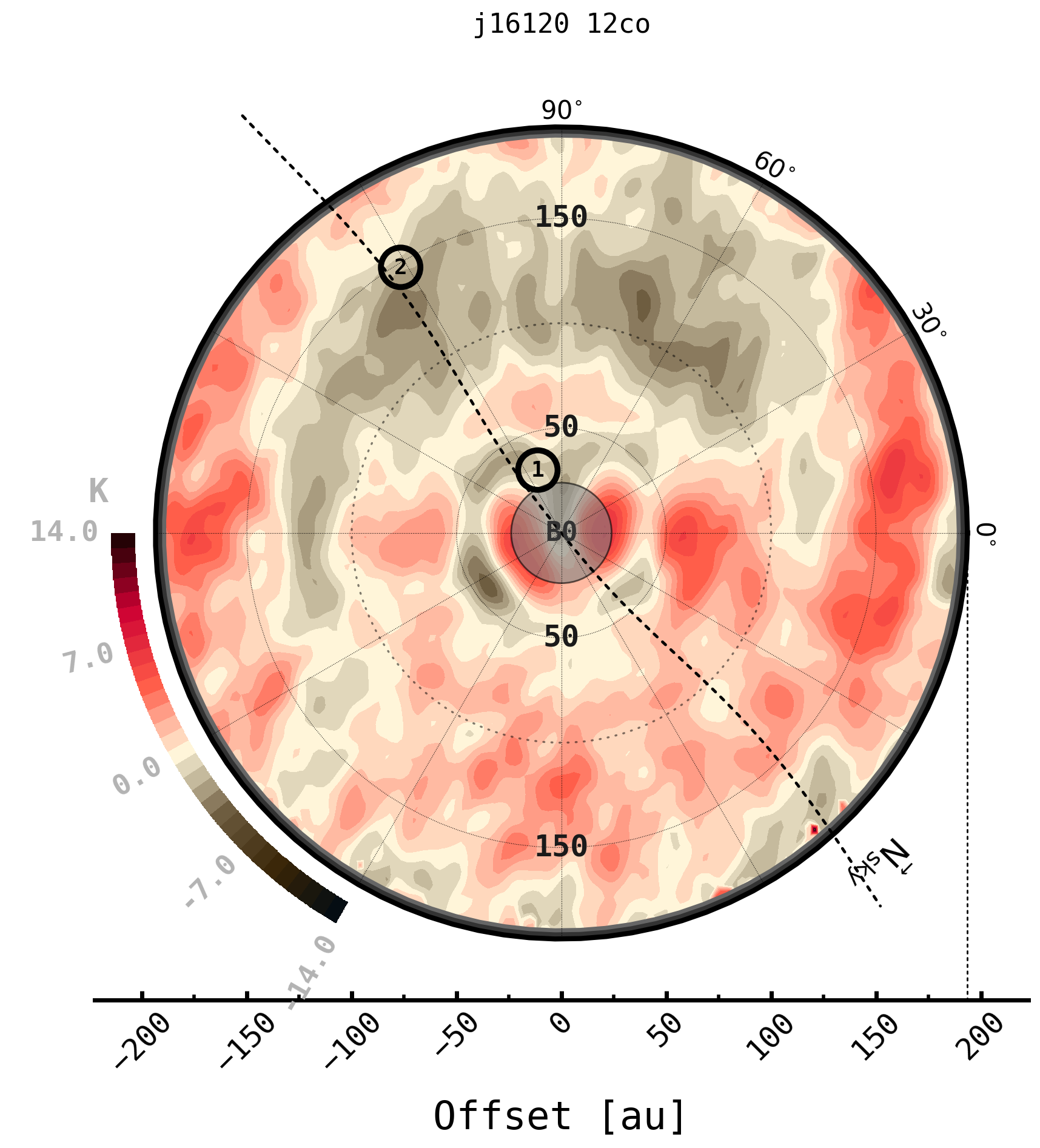}
    \caption{De-projected residual maps after subtracting the best fit \textsc{discminer} model. Left: disc velocity residuals. Middle: line width residuals. Right: peak intensity residuals. The position of the inner planet candidate at $r = 32$ au is marked by \protect\textcircled{\textbf{1}}, while that of the outer planet candidate at $r = 148$ au is marked by \protect\textcircled{\textbf{2}}. A strong positive residual, coinciding with the features observed in the individual channel maps at $\sim 50$ au, is highlighted in the left panel. Note that, in this figure, the north direction is oriented toward the bottom-right. The radial range within one beam width of the disc centre is masked in grey.}
    \label{fig:CO-residuals}
\end{figure*}

Figure \ref{fig:CO-residuals} shows the de-projected residual maps of the velocity along the line of sight (left panel), line width (middle panel), and peak intensity (right panel).
The position of the inner planet candidate suggested from ALMA Band 6 dust continuum emission \citep[32\,au, ][]{Sierra_2024} lies in the transition between a blue and red residual, similar to the spiral wakes pattern expected from a planet embedded in the disc \citep[e.g.,][]{Perez_2015, Perez_2018}. 
A strong positive residual is also observed in the vicinity of the planet candidate, at $\sim 50$ au, coinciding with the location where kink-like features were identified in the individual channel maps (Figure~\ref{fig:CO-channels}).
Additionally, positive line width residuals $\sim 0.14$\,km\,s$^{-1}$, more than five times the spectral resolution) are detected at the location of the inner planet candidate and at other positions along the same orbital radius (i.e. a ring of high velocity dispersion), similar to the expected line width residuals observed in the simulations in \cite{Dong_2019} for an embedded planet. However, there is no any particular localized peak intensity residual (e.g., tracing local heating) in the same location.

The second planet candidate proposed from the kink-like features in the outer disc \citep[originally suggested in  ][]{Sierra_2024} also lies in a region exhibiting a transition between sub-Keplerian and super-Keplerian velocities, together with positive line width residuals. However, unlike the inner planet candidate, the velocity residual pattern does not resemble that expected from a spiral wake, and the positive line width residuals do not span a broad azimuthal extent along the planet candidate's orbit.

\begin{figure*}
    \centering
    \includegraphics[width=\linewidth]{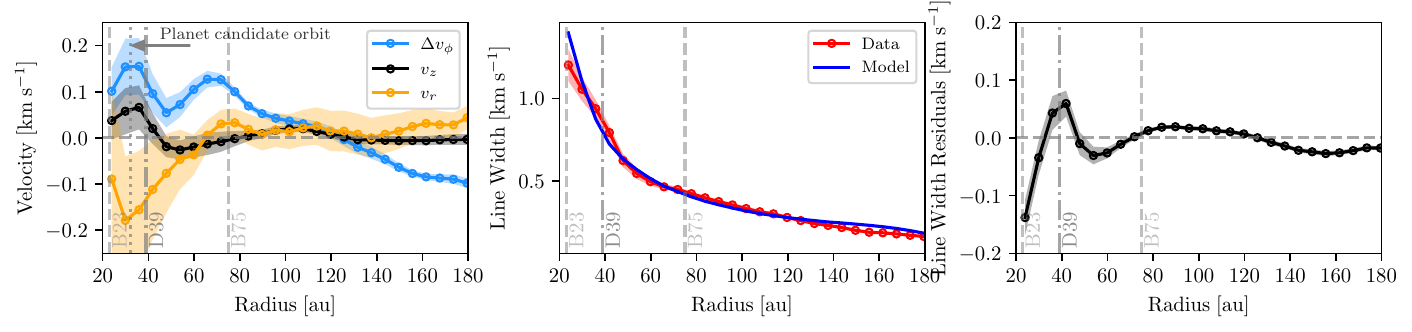}
    \caption{Left panel: Gas velocity along the vertical ($v_z$) and radial ($v_R$) axes, and the azimuthal velocity difference with respect to the azimuthal rotation model ($\Delta v_{\phi}$).
    Middle panel: Azimuthally averaged line width from the data and model. 
    Right panel: Azimuthally averaged line width residuals. The dashed/dashed-dotted vertical lines indicate the position of the rings/gaps. The dotted vertical line in the left panel indicate the radial position of the planet candidate at 32 au. The inner region within one beam size is not included.}
    \label{fig:Velocities_LineWidth}
\end{figure*}

Figure~\ref{fig:Velocities_LineWidth} presents the radial profiles of the gas velocity components, the observed and modelled line widths, and the corresponding line-width residuals derived from the $^{12}$CO \textsc{discminer} modelling. 

In the region between B23 and D39, the gas rotates at super-Keplerian velocities and their central values exhibits a radially inward flow (although the error bars are significative), while the line width simultaneously departs from the model, resulting in significant residuals. The distance between B23 and D39 is approximately one beam size; therefore, local variations within this resolution element should be interpreted with caution. We discuss these gas structures in further detail in Section~\ref{sec:Discussion_Gas}.

\subsubsection{Emission surface} \label{sec:Scale_Heights}

It is well established that different molecules trace different vertical heights \citep[e.g.,][]{Law_2021}, with optically thinner molecules probing regions closer to the disc midplane. Owing to the moderate inclination of J16120 (37.0 deg), we can study the vertical emission surface of the detected CO lines \citep[e.g.,][]{Pinte_2018b} using the non-parametric package \textsc{disksurf}, and compare it with the parametrised emission surface derived from the \textsc{discminer} modelling.

\begin{figure*}
    \centering
    \includegraphics[width=\linewidth]{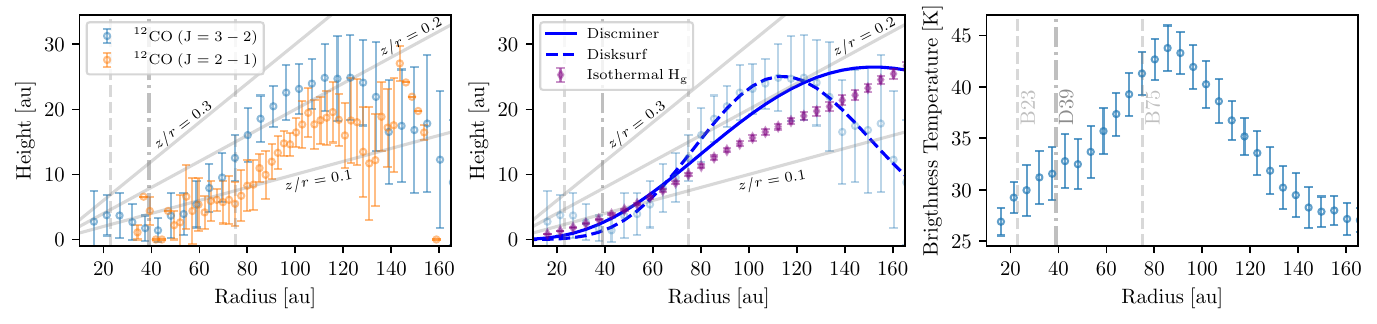}
    \caption{Left: Emission surfaces from CO isotopologue tracers and transitions. Middle: $^{12}$CO (J=3–2) \textsc{discminer} emission surface (solid line), \textsc{disksurf}  emission surface (dashed line), and isothermal scale height (dots).
    The inclined dashed lines indicate disc aspect ratios of 0.1, 0.2, and 0.3.
    Right: Radial profile of the azimuthally averaged $^{12}$CO (J=3–2) brightness temperature along the front surface.
    The vertical dashed and dashed-dotted line in all panels indicate the position of the bright and dark dust continuum rings, respectively.}
    \label{fig:Surfaces}
\end{figure*}

Left panel of Figure \ref{fig:Surfaces} shows the vertical emission profile for the $^{12}$CO (J=3--2) in this work, and $^{12}$CO (J=2--1) in \cite{Sierra_2024}. These profiles were computed by masking the data with a SNR$<$3, fixing disc geometry and offset to the values in Section \ref{sec:Dust_continuum_analysis}, and binning the profiles by a quarter of the beam size. 

The emission surface of the J=3–2 transition is located at higher altitudes than that of the J=2–1 transition by a factor of 1.3. This offset may reflect differences in optical depth, although previous studies have suggested that different transitions of the same isotopologue often trace comparable emission heights \citep[e.g.,][]{Paneque_2025, Rosotti_2025}. Around the innermost planet candidate, we find that the vertical emission surface of  $^{12}$CO (J=3--2) appears nearly flat and uniform between B23 and D39, although no significant changes are expected within this radial extent, which is comparable in size to the beam.

The middle panel of Figure \ref{fig:Surfaces} shows a comparison between the disc surfaces derived with \textsc{disksurf} and \textsc{discminer} for $^{12}$CO (J=3–2). We fit the \textsc{disksurf} data points using the same tapered power-law model adopted in \textsc{discminer}, given by
\begin{equation} \label{eq:surface}
    z(r) = z_0 \left( \frac{r}{100 \rm au} \right)^{\psi} \exp \left( - \left[\frac{r}{r_{\rm taper}} \right]^{q_{\rm taper}} \right),
\end{equation}
where $z_0$ is a characteristic height, $\psi$ is the power-law exponent, $r_{\rm taper}$ is the characteristic radius, and $q_{\rm taper}$ is the power-law of the tapered function. The best fit values are reported in Appendix \ref{app:best-fits} for $^{12}$CO (J=3--2). The fit for $^{13}$CO using \textsc{disksurf} shows strong degeneracies due to the low signal-to-noise ratio (SNR); therefore, we do not report it.
Although the best-fit values obtained from \textsc{discminer} and \textsc{disksurf} do not perfectly match, the combination of each set of parameters predicts an emission height that is consistent within the error bars. The emission surfaces only diverge beyond $\sim 120$\,au, where the geometric methodology implemented in \textsc{disksurf} tends to underestimate the emission surface in sub-Keplerian discs, as is the case for J16120 (see the left panel of Figure \ref{fig:Velocities_LineWidth}). In contrast, \textsc{discminer} appears to be less sensitive to these sub-Keplerian flows \citep[see Appendix~A of][]{Galloway_2025}.

The average elevation of the $^{12}$CO emission surface is computed following \cite{Galloway_2025}, where it is estimated by taking the mean of the binned surface profile interior to the radius where the peak surface height is found (i.e. 120 au for J16120). The resulting average elevation is $z/r = 0.15 \pm 0.07$, which is consistent with the mean value reported for discs in the Upper Scorpius star-forming region \citep{Zallio_2025}, but lower than the typical values found for bright and more extended discs \citep[$z/r > 0.3$, e.g.,][]{Law_2021, Paneque_2023, Paneque_2025, Galloway_2025}. 

The middle panel in Figure \ref{fig:Surfaces} also shows the isothermal scale height $H_{\rm g}$, which is estimated via $H_{\rm g} = c_s / \Omega_{\rm  K}$, where $c_s = \sqrt{k_B T / \mu m_H}$ is the sound speed computed from the gas temperature $T$, the Stephan-Boltzmann constant $k_B$, the mean molecular weight $\mu = 2.3$, and the Hydrogen mass $m_H$. $\Omega_{\rm K}$ is the Keplerian angular velocity for a 0.7 $M_{\odot}$ star \citep[][ and this work]{Sierra_2024}. 
The gas temperature is estimated by assuming that the $^{12}$CO emission is optically thick, and using the azimuthally averaged $^{12}$CO (J=3--2) brightness temperature profile along the front surface (right panel of Figure \ref{fig:Surfaces}).

\subsubsection{Gas mass}
\label{sec:Gas_mass}

Gas constitutes the primary mass reservoir in protoplanetary discs. Massive discs may undergo gravitational instability (GI), which can give rise to disc substructures \citep[e.g.,][]{Speedie_2024}. Therefore, constraining the gas mass of J16120 is important to assess whether GI could be responsible for its observed morphology. However, measuring the gas mass is non-trivial and typically requires chemical modelling of optically thin CO isotopologues or less abundant molecular tracers \citep{Miotello_2014, Miotello_2016}, or indirect estimates based on the dust mass. Uncertainties in the dust-to-gas mass ratio and in the relative abundances of different gas species with respect to H$_2$ make these methodologies subject to significant systematic uncertainties.

Recently, a combination of CO isotopologues and diazenylium (N$_2$H$^+$) has demonstrated to be a good tracer of the total gas mass \citep{Anderson_2019, Anderson_2022, Trapman_2022}, and it was used in the AGE-PRO large program \citep[including J16120, ][]{Zhang_2025} to derive the disc gas masses \citep{Trapman_2025}. The gas mass constrain for J16120 in the AGE-PRO collaboration is
$3.24^{+1.33}_{-0.72} \times 10^{-3} M_{\odot}$, similarly to the gas mass estimated from scaling the dust mass by a typical factor of 100 (their error bars overlap).

By modelling and comparing the obtained gas mass with the observed CO lines, \cite{Trapman_2025} determined that the relative abundance of CO with respect to H$_2$ for J16120 is $\log_{10} x_{\rm CO} = -4.82^{+0.11}_{-0.18}$, giving a CO depletion factor of $\sim 10$ compared to the ISM value of $\log x_{\rm CO} \approx -4$.

Because we only detect CO lines in Band~7, we adopt the methodology of \cite{Paneque_2025}, who demonstrated how CO depletion and the vertical location of CO emitting surfaces can be used to estimate the total disc gas mass. As mentioned above, we obtain an average aspect ratio of  $z/r = 0.15 \pm 0.07$ within $r < 120$ au (where the $^{12}$CO emission surface peaks). Therefore, assuming a Carbon depletion of 10 \citep[as constrained from ][]{Trapman_2025}, and using the scale height-disc mass relations from \cite{Paneque_2025}, extended to the lower-mass disc regime by Paneque-Carreño (in prep.), for a stellar mass of M$_\star$ = 0.6, we estimate a gas mass of M$_{\rm gas} = 0.25^{+0.60}_{-0.18}\times10^{-3}$ M$_{\odot}$. 

Note that J16120 has a higher stellar mass than the model uses. The scale height -- disc mass relationship is sensitive to stellar mass changes, with higher stellar masses yielding higher estimated disc mass values for the same aspect ratio and carbon depletion conditions, therefore our inferred disc mass from the height analysis is a lower boundary of the system, in agreement to previous constraints. Additionally, the models are full gas disc models (no substructures), however our system has a clear inner cavity, which will affect the thermal structure and inferred heights when calculating the aspect ratio from the full radial extent.

Nevertheless, the gas disc mass estimated from different methodologies is smaller than the dynamical mass of the central star by a factor of $\sim$200--2000, confirming that the disc is gravitationally stable and the observed disc morphologies are not driven by gravitational instabilities.
All values and their corresponding methodologies are summarised in Table \ref{tab:gas_mass}. 

\begin{table}
    \centering
    \caption{Gas mass estimation for J16120}    
    \begin{tabular}{c|cc}
    \hline
    Methodology &  Gas mass ($\times 10^{-3}$ M$_{\odot}$) & Reference \\
    \hline 
    CO isotopologues + N$_2$H$^{+}$     & $3.2^{+1.3}_{-0.7}$ & 1 \\
    Dust mass $\times 100$ & $2.6^{+0.2}_{-1.7}$& 1,2 \\
    CO emitting height & $0.25^{+0.60}_{-0.18}$ & This work$^{*}$ \\
    \hline
    \end{tabular}\\
    1: \cite{Trapman_2025}, 2: \cite{Sierra_2024}. $^{*}$ Following the methodology in \cite{Paneque_2025} and a Carbon depletion of\,10.
    \label{tab:gas_mass}
\end{table}

Finally, we note that the emission surface shown in Figure~\ref{fig:Surfaces} only rises beyond $\sim$D39. At smaller radii, the emission surface remains relatively flat and close to the midplane. When computing the aspect ratio relative to the radius at which the surface begins to rise ($r=39$au), we obtain $z/(r - 39\,\mathrm{au}) = 0.32 \pm 0.10$, which is approximately a factor of two higher than the aspect ratio measured relative to the central star, and significantly change the estimated disc gas mass to $4.2^{+11.9}_{-3.4} \times 10^{-3}$ M$_{\odot}$. However, although throughout this work we adopt the scale height--disc mass relation using the aspect ratio measured relative to the central star, it remains unclear how the dust and gas surface density distributions, such as the dust rings and gaps observed in J16120, affect the disc emission surface and, consequently, the inferred disc gas mass. This effect could be explored in future work using coupled gas and dust models in order to quantify their impact on the vertical disc structure and the inferred gas masses \citep[see the example on IM Lup in ][]{Deng_2025}.

\section{Discussion}\label{sec:discussion}

The morphology of the J16120 disc provides an opportunity to investigate the possible role of planet--disc interactions in shaping its observed substructures. The dust continuum emission fits well within the evolutionary sequence of planet-induced morphologies proposed by \cite{Cieza_2021, Orcajo_2025}, showing strong similarities to the fiducial stage V model. This interpretation is further supported by the relatively older age of J16120 compared to the other systems in the sequence. We therefore discuss below the main dust and gas features of the disc, with particular attention to structures that may be associated with embedded planet candidates.

\subsection{Planet Signatures in the Dust Emission?} \label{sec:Discussion_dust}

\subsubsection{No ALMA Band 7 CPD Detection}\label{sec:NoB7Detection}
A compact dust-continuum point source was reported in the ALMA Band~6 observations by \cite{Sierra_2024}. This point source was detected with an SNR of $3\sigma$ (63 $\mu$Jy), at a de-projected distance of 32 au from the central star and at a position angle of $\sim 170$ deg. The point source is not detected in the Band~7 dust-continuum observations presented in this work. Instead, diffuse continuum emission is observed within the gap (see comparison in Figure \ref{fig:B7_vs_B6}). The rms measured in a emission free region far from the disc in Bands 7 is 21~$\mu$Jy\,beam$^{-1}$ (Section~\ref{sec:observations}). However, pixel intensities within the gap are typically correlated \citep{Andrews_2021}, increasing the rms within the gap\footnote{This rms is computed by measuring the brightness standard deviation in a region within the disc gap.} to $\sigma_{\rm gap} = 98\,\mu\text{Jy\,beam}^{-1}$. Therefore, we can rule out Band 7 ($\lambda = 0.87$\,mm) point sources within the gap with a SNR $> 3\sigma_{\rm gap} \sim 300\,\mu$Jy. 

The flux observed in ALMA Band 6 in \cite{Sierra_2024} is 63\,$\mu$Jy. Therefore, assuming an optimistic optically thin spectral index of 3, the expected flux in Band 7 is $\sim 200\,\mu$Jy, which lies below $3\sigma_{\rm gap}$, making detection difficult.
For reference, the flux of the CPD around PDS~70c in Band 7 is $86 \pm 16\,\mu$Jy \citep{Benisty_2021}, which would also not be detectable at $\mathrm{SNR} > 3$ given our current rms noise level, $\sigma_{\rm gap}$. {Recently, the CPD around WISPIT2 b, which lies within a deep disc gap where the emission is negligible and the pixels are therefore not strongly correlated, was not detected either \citep{Facchini_2026}. In that work, the authors ruled out point-like emission at the location of the planet with a flux density greater than $45\,\mu$Jy at the $3\sigma$ level. These results highlight the difficulty of detecting CPDs at millimetre wavelengths.

\begin{figure}
    \centering
    \includegraphics[width=0.95\linewidth]{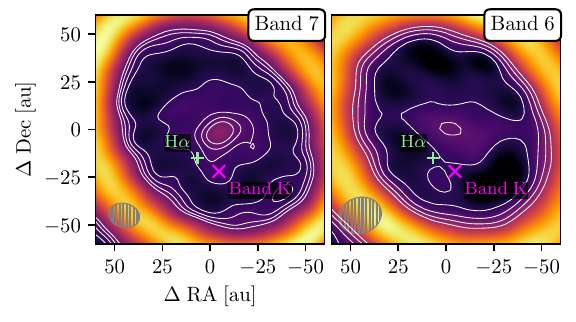}
    \caption{Central dust continuum emission from the J16120 disc observed in Band 7 (left; this work) and Band 6 \citep[right;][]{Sierra_2024}. Contours are shown at 6, 9, 12, and 15$\sigma$ for the Band 7 image, and at 3, 6, 9, and 12$\sigma$ for the Band 6 image. The $\times$ and $+$ markers are the location of the H$\alpha$ and K-band point sources in \protect\cite{Li_2025} and  \protect\cite{Ginski_2025}, respectively. The beam sizes are shown in the bottom-left corner of each panel.}
    \label{fig:B7_vs_B6}
\end{figure}

The point source in the gap of J16120 was not detected in the Band 6 dust continuum observations at higher resolution in \cite{Li_2025} either (rms of $10\mu$Jy beam$^{-1}$). However, in that work, H$\alpha$ point sources were detected with a SNR $\gtrsim\,5$ within the gap, at a de-projected distance of $\sim 23.5$\,au, and a position angle of $\sim$\,159\,deg, close to the point source detected in K band at $22.7 \pm 2.3$ au with SPHERE/IRDIS \citep{Ginski_2025}, and around the Band 6 dust continuum point source in \cite{Sierra_2024}, as shown in Figure \ref{fig:B7_vs_B6}. Their positions are still consistent within the beam size of 20.6\,au. However, \cite{Li_2025} demonstrated that the position of the H$\alpha$ point source align with the position of the point source detected in K when considering the orbital motion and the time difference between observations.

Nevertheless, in this work we find a faint dust-continuum ring at 23~au, which coincides with the de-projected radius where the H$\alpha$ emission is observed. Therefore, it is not clear how such a planet could reside on the same orbit as a dust continuum ring mainly tracing millimetre dust grains. There is no significant asymmetric feature in the B23 ring (Figure~\ref{fig:Observations}), nor a significant local intensity drop in the region around the H$\alpha$ position. However, we cannot rule out that this ring could be associated with dust trapped in the horseshoe orbit of the planet candidate \citep{Ataiee_2013, Ragusa_2017}. Deeper observations would be required to confirm or rule out this hypothesis.

Although less plausible, another possible explanation for the non-detection of a point source in the gap of J16120 in Band~7 is variability. The recent non-detection of the circumplanetary disc around PDS~70c in ALMA Band~9 \citep{Dominguez-Jamett_2025} suggests that its millimetre emission may originate from free-free emission of ionised gas shocked at the CPD surface, rather than from dust, and is therefore expected to be variable at sub-millimetre wavelengths. However, free–free emission from ionised gas is expected to be subdominant at ALMA Band~7, where the continuum is typically dominated by thermal dust emission. While limited sensitivity is likely the dominant factor behind the non-detection, the possibility that a CPD is not present cannot be ruled out.

\subsubsection{Eccentricity as a Signature of an Embedded Planet}
The dust continuum morphology of J16120 is better described by an eccentric disc (Section \ref{sec:Dust_continuum_analysis}). The best fit value for the disc eccentricity is significant ($e = 95^{+4}_{-3} \times 10^{-3}$), and may be associated with several physical processes. For example, gravitational instability (GI) may trigger non-axisymmetric structures and eccentric modes \citep{Papaloizou_1991, Laughlin_1997, Rice_2003}. However, massive discs ($M_{\rm disc} / M_{\star} > 0.1$) are needed for GI to act, which is not the case of J16120, as shown in Section \ref{sec:Gas_mass}. Stellar companions may also drive eccentricity in the central cavity, having a higher impact on the millimetre dust continuum morphology compared to that onto the gas \citep{Ragusa_2017}, and shaping the millimetre surface brightness maps \citep{Lynch_2022, Lovell_2023}.
However, using Gaia measurements, \cite{Vioque_2026} found no evidence that the astrometry of J16120 is consistent with a stellar companion within $\sim$0.1–10 au, although spectroscopic binaries could still be present at separations below 0.1 au. Likewise, a stellar companion at separations out to $\sim$10 au would likely have been detected in near-IR or optical images.

Shadows, created by a warped or misalignment inner disc, can also trigger eccentricity, due to the azimuthally varying heating, which create asymmetric pressure forces, and spiral or eccentric variation can emerge \citep[e.g.,][]{Montesinos_2016, Su_2024}, like in the disc around HD 142527 \citep{Casassus_2015, Marino_2015}. We test this hypothesis by constraining changes in the disc inclination as a function of radius using the $^{12}$CO residual velocity map (Appendix \ref{app:Wrap}). We found that the highest inclination difference is $\sim 4$\,deg. This difference is smaller than the $^{12}$CO gas opening angle estimated from the emission surface or the isothermal scale height in Section \ref{sec:Scale_Heights}, $H/r > 0.1$ (or $\sim 5.7$\,deg). Therefore, the inner disc (which is found to be highly optically thick due to its low spectral index, Figure~\ref{fig:Vis_Res_SPI}) is not expected to produce effective shadowing or a strong pressure gradient across the outer disc or ring, and is thus unlikely to drive significant eccentricity.

Eccentric substructures can also be induced by non-axisymmetric mechanisms, such as flybys \citep[e.g.,][]{Clarke_1993, Cuello_2019}, or the Rossby Wave Instability \citep[e.g.,][]{Pierens_2018}. However, there is currently no evidence for either in J16120. 

As such, the leading explanation for the observed eccentricity is the presence of a planet embedded within the disc dust-gap. Several works \cite[e.g.,][]{Kley_2006, Ataiee_2013, Pinilla_2015b, Teyssandier_2017, Ragusa_2018} have demonstrated that a planet to star mass ratios larger than $3 \times 10^{-3}$ can lead to eccentricity growth in the disc. For J16120, with a central star of $0.7$\,M$_{\odot}$ star, the planet candidate should have a mass M$_{\rm p} \gtrsim 700$\,M$_{\oplus} = 2.2$ M$_{\rm Jup}$, consistent with the 4 M$_{\rm Jup}$ planet mass constrained from the SPHERE band K and H$\alpha$ emission \citep{Li_2025}.

Recent studies have shown that disc eccentricity is expected to decrease rapidly beyond the planet orbit \citep{Padgett_2026}. In the case of J16120, the eccentricity may therefore be largely confined to the inner disc and the B23 ring, both of which lie within the orbit of the 32 au planet candidate. This scenario would also explain why the eccentricity does not strongly affect the large-scale disc kinematics \citep[e.g., as those observed in MWC 758][]{Izquierdo_2026b}, where eccentric discs are expected to produce elongated structures \citep{Ragusa_2024}.

\subsection{Signatures of a Planet in the Gas Emission}
\label{sec:Discussion_Gas}

A large diversity of gas substructures is observed and inferred in J16120. The channel maps of $^{12}$CO and $^{13}$CO (Figure \ref{fig:CO-channels}) show a kink like structure in the South of the disc, in a de-projected orbit between $\sim$ 125 and 171 au, and over a large range of velocities  (from $\sim$ 3.7 to 4.5 km s$^{-1}$ ), indicating possible extended deviations from Keplerian motion rather than localised substructures. 
Additionally, the azimuthal position of the feature is not the same through different channel maps, but it moves following approximately the same orbit, which is between the de-projected $R_{90}$ radius of $^{12}$CO and $^{13}$CO. Similar features are also observed in some channel maps on the opposite side of the disc (towards the north), which could suggest the presence of a gas gap. However, we do not detect a clear gap at these radii in either the moment 0 or peak intensity maps, nor in the radial profiles shown in Figure~\ref{fig:Observations}. Consequently, we consider the gas-gap interpretation as the origin of the features to be inconclusive.

Similar structures have been also observed in other discs around the outer border of the disc \citep[e.g.,][]{Paneque_2021},  where it has been interpreted as projected emission from the upper and lower side of the disc, which is connected by gas around the disc midplane. An inspection of the Keplerian \textsc{discminer} model (top panels of Figure~\ref{fig:CO-channels_Model}) shows that the combined front- and backside emission can reproduce these structures, resulting in negligible residuals (top-middle panels of Figure~\ref{fig:CO-channels_Model}). Therefore, the superposition of front- and backside emission mimics the kink-like structures observed in the outer disc, confirming previous interpretations of these substructures and suggesting that the gas features detected beyond the dust continuum emission in this work and by \cite{Sierra_2024} may not have a planetary origin.

On the other hand, the inner planet candidate at 32\,au presents new interesting evidence that suggest a planetary origin. The \textsc{discminer} residuals maps in Figure \ref{fig:CO-residuals} around the inner planet candidate presents 1) a localised velocity residual similar to a Doppler flip, and 2) a localised positive line width residual at the location of the planet and along its orbit. 
Additionally, the $^{12}$CO channel maps in the region around 50 au (just exterior to the inner planet's orbit) exhibit asymmetries that resemble kink-like features, and the gas kinematics also suggests radially inward flow towards the disc cavity. All these properties may suggest a planetary origin and are individually discussed below.

\subsubsection{Planet Kinematic Signatures in the Velocity Residual Map}
The localized velocity residual around the inner planet candidate is associated with a spiral wake-like structure, which changes from sub-Keplerian to super-Keplerian velocities across the location of the planet candidate. This pattern may be associated with a Doppler flip caused by the presence of a planet \citep[e.g.,][]{Casassus_2019}, supporting the hypothesis of a planet at this location, or changes in the disc geometry \citep[e.g.,][]{Calcino_2024}.

We also observe kink-like features at $\sim 50$ au (Figure~\ref{fig:CO-channels}), exterior to the planet's orbit at 32 au. The velocity residuals in the moment maps (left panel of Figure~\ref{fig:CO-residuals}) reveal a localised positive residual at $\sim 50$ au. This residual lies within one beam of the planet candidate's orbit and may be related to the kinematic signatures expected from the planet.

As mentioned above, some of the large-scale residual velocities may also be associated to changes in the disc geometry, and in particular to a warped disc, as recently shown for the exo-ALMA discs \citep{Teague_2025, Winter_2025}. 
We use the methodology and tools in \cite{Winter_2025} to study the velocity residual maps in our work, and look for radial variations of the disc inclination and position angle. The results (shown in Appendix \ref{app:Wrap}) show a maximum inclination variation of $\sim 4$\,deg, and a maximum PA variation of $\sim 3$\,deg, a bit lower than the variation for MWC 758 \citep{Winter_2025}. The best-fitting model reproduces most of the observed residuals at $r < 60$ au, and the positive residuals on the western side of the disc (left panel of Figure~\ref{fig:Warp}). However, it fails in reproducing the positive localised velocity residual in the vicinity of the planet candidate at $\sim 50$ au (marked with an arrow in Figure \ref{fig:Warp}), as revealed by the Residual map (top-right panel in Figure \ref{fig:Warp}). The origin of the latter remains uncertain. However, as it lies in the outer vicinity of previously identified tracers suggesting the presence of a planet candidate, it may likewise be induced by the same planetary companion.

At the location of the outer planet candidate, the velocity residuals also transition from sub- to super-Keplerian velocities. However, unlike the inner planet candidate, they are not connected to a spiral wake-like structure, making this outer feature less compelling as evidence for a planet, while its origin remains uncertain.

\subsubsection{Line width residuals revealing an embedded planet}

The line-width residual map in the middle panel of Figure \ref{fig:CO-residuals} shows distinct substructures. The positive residuals near the inner planet candidate and along its orbit resemble the signatures predicted by the hydrodynamic simulations of \cite{Dong_2019}, where a planet is embedded in a protoplanetary disc in a circular orbit of 30 au and around a 1 $M_{\odot}$ star, similar to the position of the proposed inner planet ($\sim$32 au) and stellar parameters of J16120. The origin of the enhanced line width at the planet's orbit in their simulations is increased velocity dispersion due to planet-induced turbulence.

\begin{figure*}
    \centering
    \includegraphics[width=\linewidth]{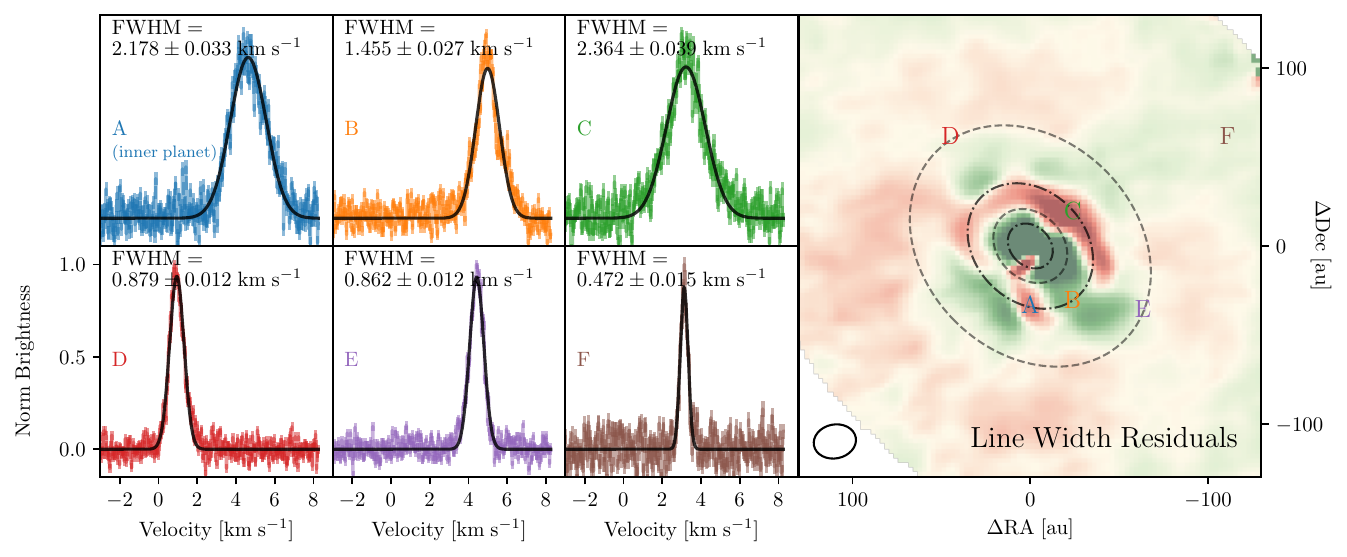}
    \caption{$^{12}$CO line profiles extracted at selected pixel locations; the corresponding coordinates are over-plotted on the line width residual map on the right. The dashed ellipses indicate the position of the B23, B75 rings, while the dashed-dotted are the D14, D39 rings. The Full-Width-Half-Maxima of each line profile is reported in the top-left corner of each panel. Pixel A is located at the position of the inner planet candidate at 32 au.}
    \label{fig:LineWidths}
\end{figure*}

The azimutally-averaged line width radial profile of J16120 deviates from the model by $\sim 0.05$ km s$^{-1}$ (6\%) at D39 (Figure \ref{fig:Velocities_LineWidth}). 
Such a deviation is much smaller than the line-width broadening expected for a 4~$M_{\rm Jup}$ planet, and slightly smaller than the 10\% non-thermal velocity dispersion predicted for a 1~$M_{\rm Jup}$ planet in the simulations of \cite{Dong_2019}, which assume a fixed disc viscosity of $\alpha = 10^{-3}$ and a vertically isothermal disc structure. This suggests that, if the enhanced line broadening observed in J16120 is produced solely by an embedded planet, and the disc parameters of J16120 are similar to those in \cite{Dong_2019}, the mass of the planet is likely constrained to $\sim 1~M_{\rm Jup}$.
However, it is important to note that spatial smearing can artificially reduce the observed velocity dispersion, as demonstrated by \cite{Izquierdo_2026b}, implying that the inferred planet mass may be underestimated and that the estimated value of $1\,M_{\rm Jup}$ should therefore be regarded as an approximation to lower limit mass.

On the other hand, positive line width residuals are also present at the location of the outer planet candidate. However, the positive residual structure along this orbit extends over only $\sim$ 70 deg in azimuth and is weaker than that observed for the inner planet candidate, making these features insufficiently robust to be considered reliable signatures of an embedded planet.

Figure \ref{fig:LineWidths} shows the line profile at different disc locations. Positions A, B, and C lie at the same orbital radius as D39. At position A and C, where the line width residuals presents an excess, the full-width-half-maxima (FWHM) is 2.178 and 2.364 km s$^{-1}$, respectively. These values are a factor of 1.5 and 1.6 larger than the FWHM at B, where the line width is narrower than the average value in the same orbit.

The line widths at position D and E (same orbital radius as B75) show similar FWHM of $\sim\,0.8$\,km s$^{-1}$, as also expected from the line width model\footnote{Note that the line width from the \textsc{discminer} model is parametrised using the half width of the profile at half power.} (middle panel of Figure \ref{fig:Velocities_LineWidth}). At position F (far from the disc centre $\sim 155$ au), the FWHM decreases to 0.472 km s$^{-1}$.
No double peak is observed in the spectra at the locations where the line width is enhanced, possibly due to the difficulty to distinguish between the front side and back side of the disc at  39 au.

The similarity of the line width residuals in J16120 compared with the simulations in \cite{Dong_2019} is landmark observational case to date. The coincidence of the line width residuals with previously reported planet-formation signatures from different tracers \citep{Sierra_2024, Ginski_2025, Li_2025} around the same region is suggestive of a forming planet in this orbit. However, a comprehensive understanding of the line width enhancement requires exploring the role of additional physical processes beyond a planetary origin.
Future studies combining multi-line observations (e.g., SO tracing local planet heating), improved thermochemical modelling, and self-consistent MHD simulations will be key to isolating the specific contribution of an embedded planet from other mechanisms that may also contribute to line broadening. For example, the simulations in \cite{Barraza-Alfaro_2025} clearly identify and separate the effects of the vertical shear instability, the magnetorotational instability, and the gravitational instability in the localised kinematic signatures observed in the exoALMA Large Program \citep{Teague_2025}. A similar effort is required to disentangle their respective contributions to line-width enhancement.

\subsubsection{Hints of Gas Flow Towards the Planet's Orbit}

Gas kinematics can also be used to search for evidence of planet formation in gas-rich protoplanetary discs. For example, the presence of forming planets may trigger meridional flows \citep[e.g.,][]{Moorbidelli_2014} that transport material from the disc midplane—where the dynamics are dominated by planetary torques—to the disc surface, which is dominated by viscous torques \citep[e.g.,][]{Dong_2019, Kley_2001}. Radially, gas in the midplane is driven away from the planet’s orbit by Lindblad torques \citep[e.g.,][]{Masset_2002}, while in the disc surface layers it moves toward the orbit due to the pressure drop induced by the planet. For optically thick line emission, the observed kinematics predominantly trace the disc surface, where the gas is therefore expected to flow toward the planet’s orbit. 

In the specific case of J16120, the left panel of Figure~\ref{fig:Velocities_LineWidth} shows a trend towards negative radial velocities ($v_r$) beginning at B75 and reaching a minimum at 32 au, corresponding to the location of the inner planet candidate. Interior to the planet's orbit, we have only a single radial measurement, limiting our ability to constrain the radial velocity in that region.
Note that the velocity uncertainties remain significant for $r < 75$ au and are also consistent with zero, so these measurements should be interpreted with particular care. In the outer disc (beyond B75), the radial velocity is flat and around zero.

\begin{figure}
    \centering
    \includegraphics[width=\linewidth]{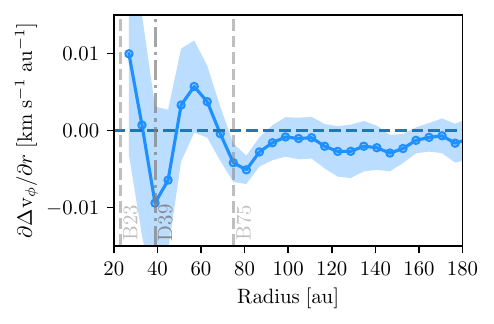}
    \caption{Derivative of the azimuthal velocity deviation relative to the Keplerian disc. The vertical dashed and dashed-dotted line indicate the position of the bright and dark dust continuum rings, respectively.}
    \label{fig:Derivative_vphi}
\end{figure}

On the other hand, the gas is super-Keplerian inside 120 au ($\Delta v_{\phi} > 0$), including the vicinity of the inner planet candidate. The radial gradient of $\Delta v_{\phi}$ becomes negative around B75, reaching a local minimum at approximately the same location, as shown in Figure~\ref{fig:Derivative_vphi}. Co-spatiality between millimetre rings and pressure maxima has been observed in many discs \citep{Rosotti_2020, Izquierdo_2023, Stadler_2025}. This behaviour is indicative of dust trapping and was previously suggested for J16120 based on Band 6 observations \citep{Sierra_2024}.

Finally, the vertical velocity ($v_z$) tends toward positive values around the planet candidate’s orbit; however, the associated uncertainties and limited angular resolution make it difficult to draw firm conclusions about the presence of meridional flows in this region, such as those observed in HD\,163296 \citep{Teague_2019Nature}.

\section{A planet embedded in the J16120 disc?}\label{sec:Hints}

Multiple lines of evidence for a forming planet have been reported within the gap of J16120, including a CPD candidate identified in ALMA Band 6 observations \citep{Sierra_2024}, point sources detected in the H and K bands with SPHERE \citep{Ginski_2025}, H$\alpha$ emission \citep{Li_2025}, line-width enhancements tracing increased velocity dispersion around the planet candidate and along its orbit, as well as a suggestive Doppler flip and gas flow directed towards the planet's orbit (this work). Although the exact radial locations of these tracers differ, they remain consistent within the observational uncertainties, as shown in Figure \ref{fig:Hints}. Moreover, some of the apparent offsets may arise from orbital motion. For example, \cite{Li_2025} showed that the positional offset between the K-band and H$\alpha$ detections can be explained by the orbital motion of the planet candidate over the time elapsed between the observations. 

\begin{figure*}
    \centering 
    \includegraphics[width=0.7\linewidth]{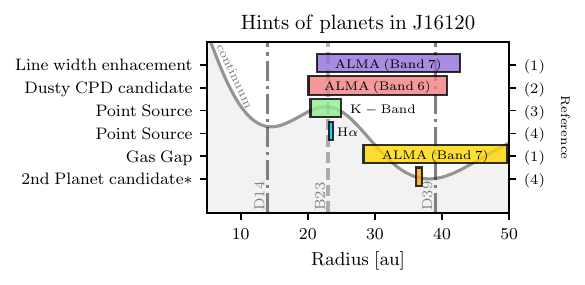}
    \caption{Summary of the different signatures of planet formation within the gap of J16120. The background profile shows the radial dust continuum emission observed in Band 7 (this work). The vertical lines indicate the position of rings (B) and gaps (D).
    References: (1) This work, (2) \protect\cite{Sierra_2024}, (3) \protect\cite{Ginski_2025}, (4) \protect\cite{Li_2025}. *: Second, as-yet undetected, planet candidate proposed by \protect\cite{Li_2025}.}
    \label{fig:Hints}
\end{figure*}

In Figure \ref{fig:Hints}, we also include the radial extent of the gas gap observed in the current $^{13}$CO data, as well as a second, as-yet undetected planet candidate proposed by \cite{Li_2025} at an orbital radius of $\sim 37$ au based on the gap size and the massive accreting gap-planet model of \cite{Close_2020}. Interestingly, this orbital radius coincides with the location of the D39 gap reported in this work. Such an additional planet candidate could also play a significant role in shaping the disc eccentricity \citep[e.g.,][]{Lee_2002}, such as the eccentric structure described in Section \ref{sec:Galario}.

Future observations will be crucial for testing the planet formation scenario in J16120. In particular, VLTI/GRAVITY+ spectroscopy, infrared observations with the James Webb Space Telescope, millimetre observations targeting gas tracers of planet formation, and new high-resolution ALMA data will provide complementary constraints on the nature of the proposed planet candidate. These observations will help to characterise the circumplanetary environment, probe ongoing accretion processes, and search for additional kinematic and structural signatures of planet--disc interactions. Ultimately, these data will help determine whether the observed substructures can be robustly attributed to ongoing planet formation within the gap of J16120.

\section{Conclusions}\label{sec:conclusions}
We present new ALMA Band~7 observations of the disc around J16120, revealing additional evidence for planet formation associated with a candidate embedded within the dust-disc gap. These results provide further support for the presence of a planetary companion previously suggested by millimetre \citep{Sierra_2024}, infrared \citep{Ginski_2025}, and H$\alpha$ observations \citep{Li_2025}. We summarise our main findings below:

\begin{itemize}
    \item The detected dust continuum, $^{12}$CO (J=3--2), $^{13}$CO (J=3--2) emission reveal a bright disc and a ring-like structure peaking at 75 au, consistent with morphologies previously observed at millimetre and infrared wavelengths.  
    The Band~6 and Band~7 dust continuum data also reveal a faint ring at 23 au and gaps at 14 and 39 au. The spectral index between the two bands exhibits local maxima at the gap locations (D14 and D39) and local minima at the rings (B23 and B75), suggesting the presence of dust traps. This interpretation is further supported by local minima in the radial derivative of the azimuthal gas velocity, which are also indicative of dust trapping. No ALMA Band 7 ($\lambda = 0.87$\,mm) point sources are detected within the  disc gap at a SNR > 3$\sigma_{\rm gap} \sim$ 300 $\mu$Jy.
    
    \item The residual map of the Band~7 ($\lambda = 0.87$\,mm) dust continuum emission, obtained after subtracting the azimuthally averaged profile, shows positive residuals that align with the $m=1$ spiral structure also observed in Band~6 ($\lambda = 1.3$\,mm). However, most of these features disappear when the dust continuum morphology is modelled with an eccentric disc rather than an axisymmetric disc. The best-fitting eccentric disc model yields an eccentricity of $e = 95^{+4}_{-3} \times 10^{-3}$, and we suggests that its origin may be associated with the presence of a 
    planetary-mass companion (with a mass of $\gtrsim 2.2$ M$_{\rm Jup}$)  embedded within the disc gap, consistent with previous constraints obtained from SPHERE band K and H$\alpha$ emission.

    \item The gas mass of J16120, estimated from the $^{12}$CO emission surface following the methodology in \cite{Paneque_2025}, is $0.25 ^{+0.60}_{-0.18} \times10^{-3}\,M_{\odot}$. This is value is lower than previous estimates based on CO isotopologue and N$_2$H$^{+}$ observations \citep{Trapman_2025}; however, it should be interpreted as a lower limit on the disc gas mass. For all estimated gas masses, the disc is insufficiently massive to become gravitationally unstable and account for the observed disc substructures.
    
    \item After subtracting a non-eccentric Keplerian model, the $^{12}$CO velocity residual map reveals spiral wake-like structures that appear to originate from a region near the location of the inner planet candidate at 32 au. This planet candidate lies at the transition between sub-Keplerian and super-Keplerian rotation, resembling the Doppler flip expected from an embedded planet. 
    
    \item The residual map of the $^{12}$CO line width shows a prominent positive residual (amplitude of $\sim 0.14$ km\,s$^{-1}$) at the location of the planet candidate at 32 au, as well as similar positive residuals along the same orbit. These features are consistent with the expected signatures of an embedded planet \citep{Dong_2019}, which enhances the local velocity dispersion along its orbit. This suggests that a planetary companion may be responsible for the observed line-width broadening in J16120. Based on the simulations of \cite{Dong_2019} and accounting for the spatial smearing of the observations \citep{Izquierdo_2026b}, the planet mass is constrained to be $\gtrsim 1\,M_{\rm Jup}$.
    
    \item Hints of gas flowing toward the orbital radius of the planet candidate are inferred from the disc kinematics, although the associated error bars remain significant. Such a flow is nevertheless expected from the pressure minima within a gap carved by an embedded planet. Similarly, the vertical gas velocity appears to suggest a flow from the disc midplane toward the upper disc layers at the location of the planet candidate. However, these interpretations should be treated with caution, as the uncertainties associated with the velocity component remain substantial, preventing robust conclusions from being drawn.

    \item We did not detect any robust signatures of an outer planet beyond the dust continuum emission. However, we confirm the presence of a feature that resembles a kink-like structure \citep[previously reported in Band 6 by][]{Sierra_2024}, located at an orbital radius between 125 and 171\,au. Given its large azimuthal extent, its proximity to the outer edge of the gas disc, and the fact that gas disc modelling indicates consistency with Keplerian motion (residuals below 5$\sigma$), we suggest that this feature is unlikely to be of planetary origin and may instead be associated with projected emission from the upper and lower surfaces of the disc.
    
\end{itemize}

The dust continuum eccentricity, localised velocity structures, enhanced line widths, and disc kinematics provide new and compelling evidence for a planetary companion at an orbital radius of $\sim$32 au in J16120. Together with previous hints of planet formation, these findings make J16120 one of the strongest candidates for hosting an embedded protoplanet, alongside the confirmed planetary systems PDS\,70 and WISPIT\,2.  Future observations at shorter and longer millimetre wavelengths, in both dust continuum and molecular lines, particularly those sensitive to local planet-induced heating like SO, SiS, HCN, or C$_2$H \citep[][]{Law_2023, Izquierdo_2026}, will be essential to confirm these findings and to further characterise the planet embedded within the gap of J16120.

\section*{Acknowledgments}
This paper makes use of the following ALMA data: ADS/JAO.ALMA\#2023.1.01100.S ALMA is a partnership of ESO (representing its member states), NSF (USA) and NINS (Japan), together with NRC (Canada), NSTC and ASIAA (Taiwan), and KASI (Republic of Korea), in cooperation with the Republic of Chile. The Joint ALMA Observatory is operated by ESO, AUI/NRAO and NAOJ. The National Radio Astronomy Observatory is a facility of the National Science Foundation operated under cooperative agreement by Associated Universities, Inc.
This research was supported in part by grant NSF PHY-2309135 to the Kavli Institute for Theoretical Physics (KITP).

A.S. acknowledges support from UNAM DGAPA-PAPIIT grant IG101224.
A.S. and P.P. acknowledge funding from the UK Research and Innovation (UKRI) under the UK government’s Horizon Europe funding guarantee from ERC (under grant agreement No 101076489).
T.P.-C. acknowledges support from the Michigan Society of Fellows and has received funding by the Heising-Simons foundation through the 51 Pegasi B Fellowship.
L.P. acknowledges support from ANID BASAL project FB210003 and ANID FONDECYT Regular 1262272.
C.A.G. acknowledges support from Comité Mixto ESO-Gobierno de Chile 2023, under grant 072-2023. 
A.R. has received funding from the Royal Society through a University Research Fellowship grant number URF\textbackslash R1\textbackslash 241791.
L.A.C. acknowledges support from the Millennium Nucleus on Young Exoplanets and their Moons (YEMS), NCN2024\_001, Chile and FONDECYT Regular grant 1241056.
S.F. acknowledges financial contributions from the European Union (ERC, UNVEIL, 101076613) and from PRIN-MUR 2022YP5ACE. Views and opinions expressed are however those of the authors only and do not necessarily reflect those of the European Union or the European Research Council. Neither the European Union nor the granting authority can be held responsible for them.
C.F.M. is Funded by the European Union (ERC, WANDA, 101039452). Views and opinions expressed are however those of the author(s) only and do not necessarily reflect those of the European Union or the European Research Council Executive Agency. Neither the European Union nor the granting authority can be held responsible for them.
J.M. acknowledges support from ANID -- Millennium Science Initiative Program -- Center Code NCN2024\_001.
G.R. acknowledges support from the European Union (ERC Starting Grant DiscEvol, project number 101039651) and from Fondazione Cariplo, grant No. 2022-1217. Views and opinions expressed are, however, those of the author(s) only and do not necessarily reflect those of the European Union or the European Research Council. Neither the European Union nor the granting authority can be held responsible for them. 
\section*{Software}
This work made use of the following software: \textsc{Astropy} \citep{astropy:2018}, \textsc{CASA} \citep{McMullin_2007}, \textsc{Discminer} \citep{Izquierdo_2021}, \textsc{Emcee} \citep{Foreman_2013}, \textsc{Frankenstein} \citep{Jennings_2020}, \textsc{Matplotlib} \citep{Matplotlib_2007}, \textsc{Numpy} \citep{Numpy_2020}.

\section*{Data availability}
The self-calibrated data underlying this article will be available at \href{https://zenodo.org/}{zenodo.org}. The non-calibrated data is publicly available at \href{https://almascience.nrao.edu/aq/}{almascience.nrao.edu/aq} using the project code 2023.1.01100.S.



\bibliographystyle{mnras}
\bibliography{main} 



\appendix

\section{Background millimetre object} \label{sec:background}

We report the Band 7 ($\lambda = 0.87$\,mm) detection of a resolved millimetre source at a projected distance of $\sim 4.4^{\prime \prime}$ ($\sim 582$ au) from the J16120 centre (Figure \ref{fig:back}). The dust continuum flux of this source is $0.81 \pm 0.02$ mJy, and no $^{12}$CO nor $^{13}$CO emission is detected in its surroundings. This source was also observed in Band 6 in \cite{Sierra_2024}, with a dust continuum flux of $0.12 \pm 0.02$ mJy. Therefore, its spectral index between ALMA Band 7 ($\lambda = 0.87$\,mm) and 6 ($\lambda = 1.3$\,mm) is $\alpha = 4.56 \pm 0.40$, suggesting optically thin emission from small dust grains. However, the steep spectral index, lack of associated CO emission, and large projected separation also make a background sub-millimetre galaxy the most plausible interpretation.

\begin{figure}
    \centering
    \includegraphics[width=\linewidth]{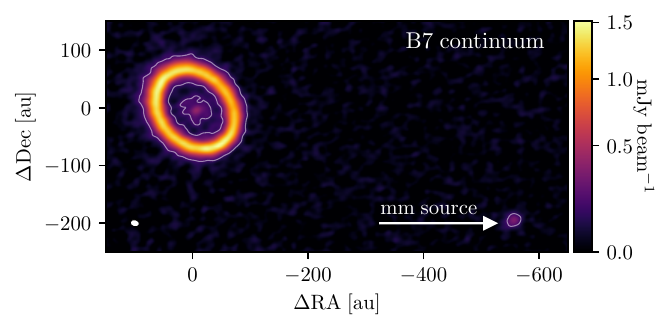}
    \caption{Extended Band~7 dust continuum emission, within which a sub-millimetre source, potentially a background sub-millimetre galaxy, is detected. Contours indicate the ALMA Band~7 dust continuum emission at the 7-$\sigma$ level.}
    \label{fig:back}
\end{figure}

\section{Best-fit Model Parameters}
\label{app:best-fits}

Table \ref{tab:Eccentricity} summarises the best-fitting parameters of the eccentric disc model presented in Section \ref{sec:Galario}.
Table \ref{tab:Discminer} presents the best fit model parameters for the \textsc{discminer} model in both $^{12}$CO (J=3--2) and $^{13}$CO (J=3--2). The parameters presented in the table corresponds to the parametrization of the attributes as following. 
\begin{itemize}
    \item Peak intensity: $I_{p} = I_{p0} (R/100 {\rm au})^{p_p} (z/100 {\rm au})^{q_p}$
    \item Line width: $L_w = L_{w0} (R/100 {\rm au})^{p_w} (z/100 {\rm au})^{q_w}$    
    \item Line slope: $L_s = L_{s0} (R/100 {\rm au})^{p_s}$   
    \item Upper surface: $z_U = z_{U0} (R/100 {\rm au})^{p_U} \exp(-(R/R_{Ut})^{q_U})$
    \item Lower surface: $z_L = z_{L0} (R/100 {\rm au})^{p_L} \exp(-(R/R_{Lt})^{q_L})$
\end{itemize}
Additionally, the mass of the central star (M$_\star$), the systemic velocity (v$_{\rm sys}$), the disc inclination, and position angle are free parameters of the fit. See \cite{Izquierdo_2021} for more details.

Table \ref{tab:surface} lists the best-fitting emission-surface parameters derived with \textsc{discminer} for $^{12}$CO (J=3--2) and $^{13}$CO (J=3--2), and with \textsc{disksurf} for $^{12}$CO (J=3--2), as described in Section \ref{sec:Scale_Heights}. The error bars from the \textsc{discminer} fit in Table \ref{tab:Discminer} and \ref{tab:surface} are derived from the walker distributions by computing the 16th and 84th percentiles. However, these uncertainties are likely underestimated. Using spatial correlations in the image plane, \citet{Hilder_2025} showed that accurately quantifying the true errors is not straightforward, but that they may be up to an order of magnitude larger than those inferred from the Bayesian analysis.

\begin{table}
\centering
\caption{Best fit parameters of the eccentric disc modelling.}
\begin{tabular}{ccc}
\hline
Parameter  [Units] & Value \\
\hline
$\Delta$RA     [mas] & $87.6^{+2.5}_{-1.9}$ \\
$\Delta$Dec    [mas] & $34.6^{+2.2}_{-1.6}$ \\
$e$                  & $95^{+4}_{-3} \times 10^{-3}$ \\
$\omega$       [deg] & $120 \pm 2$ \\
$r_0$          [au]  & $4.3^{+1.2}_{-4.3}$ \\
$r_{23}$       [au]  & $22.5^{+2.5}_{-4.0}$ \\
$r_{75}$       [au]  & $76.4 \pm 0.1$ \\

$\sigma_0$     [au]  & $1.5^{+3.1}_{-0.2}$ \\
$\sigma_{23}$  [au]  & $10.9^{+3.3}_{-2.1}$ \\
$\sigma_{75}$  [au]  & $9.2 \pm 0.1$ \\

$A_0$  [$\mu$Jy/pix]   & $8.5^{+2.3}_{-3.3}$ \\
$A_{23}$ [$\mu$Jy/pix] & $1.2 \pm 0.1$ \\
$A_{75}$ [$\mu$Jy/pix] & $12.1 \pm 0.1$ \\
\hline
\end{tabular}
\label{tab:Eccentricity}
\end{table}

\begin{table}
    \caption{Best fit parameters for the \textsc{discminer} model.}
    \centering
    \begin{tabular}{c|cc}
    \hline
    Parameter [Units] & Best-fit ($^{12}$CO) & Best-fit ($^{13}$CO)  \\
    \hline
    M$_\star$ [M$_{\odot}$]     &  $0.7 \pm 0.1$   & $0.7 \pm 0.1$   \\
    v$_{\rm sys}$ [km s$^{-1}$] &  $4.527 \pm 0.025$   & $4.524 \pm 0.025$   \\

    i [deg]& $38.3^{+0.1}_{-0.1}$ & 37.0* \\
    PA [deg] & $44.4^{+0.1}_{-0.1}$ & 45.1* \\
    
    $I_{p0}$ [mJy pix$^{-1}$]   &  $0.37 \pm 0.01$ & $0.55 \pm 0.01$ \\
    $p_p$                       &  $-3.26 \pm 0.01$ & $-0.53 \pm 0.06$ \\
    $q_p$                       &  $1.80 \pm 0.01$  & $1.13 \pm 0.04$\\

    $L_{w0}$ [km s$^{-1}$]   & $0.54 \pm 0.01$  & $1.45 \pm 0.23$ \\
    $p_w$                    & $-1.20 \pm 0.01$ & $-0.32 \pm 0.03$\\
    $q_w$                    & $0.28 \pm 0.01$  & $0.49 \pm 0.03$\\

    $L_{s0}$                 & $1.93 \pm 0.01$ & $1.66 \pm 0.02$\\
    $p_s$                    & $0.35 \pm 0.01$ & $0.67 \pm 0.03$\\
    
    $z_{U0}$ [au]  &  $44.6 \pm 0.6$ & $25.3 \pm 1.6$  \\
    $p_U$          &  $2.8 \pm 0.1$ & $2.9 \pm 0.1$ \\
    $q_U$          &  $1.7 \pm 0.1$ & $1.4 \pm 0.1$ \\
    $R_{Ut}$ [au]  & $110.0 \pm 1.0$  & $58.2 \pm 1.7$ \\

    $z_{L0}$ [au] &  $22.1 \pm 0.1$  & $1.1 \pm 0.2$   \\
    $p_L$         &  $1.84 \pm 0.01$ & $0.28 \pm 0.07$ \\
    $q_L$         &  $1.06 \pm 0.01$ & $3.03 \pm 0.80$ \\
    $R_{Lt}$ [au] &  $228.2 \pm 2.0$ & $388.3 \pm 44.5$\\
    
    \hline         
    \end{tabular}
    \newline
    *: The disc inclination and position angle for $^{13}$CO were fixed to those computed from the dust continuum (Section \ref{sec:Dust_continuum_analysis}). 
    \label{tab:Discminer}
\end{table}

\begin{table}
    \centering
    \caption{Best fit parameters of the upper emission surface for the molecular lines in this work.}  
    \begin{tabular}{c|ccc}
    \hline
     \multirow{2}{*}{Parameter} & \textsc{Disksurf} & \textsc{Discminer}  & \textsc{Discminer} \\
       & $^{12}$CO (J=3--2) & $^{12}$CO (J=3--2) & $^{13}$CO (J=3--2) \\
     \hline
     $z_0$ [au]           & $93.5^{+56.2}_{-38.4}$ & $44.6 \pm 0.6$  & $ 25.3 \pm 1.6$ \\
     $r_{\rm taper}$ [au] & $86.6^{+18.5}_{-13.8}$ & $110.0 \pm 1.0$ & $58.2 \pm 1.7$ \\     
     $\psi$               & $4.4 \pm 0.7$    & $2.8 \pm 0.1$   & $2.9 \pm 0.1$ \\
     $q_{\rm taper}$      & $2.3^{+0.6}_{-0.4}$    & $1.7 \pm 0.1$   & $1.4 \pm 0.1$\\
     \hline
    \end{tabular}
    \label{tab:surface}
\end{table}

\section{More about Discminer modelling}
\label{app:Discminer}
Figure~\ref{fig:CO-channels_Model} shows the \textsc{discminer} models of the $^{12}$CO and $^{13}$CO emission (top panels) and the corresponding residuals (bottom panels), displayed for the same channel maps as the data shown in Figure \ref{fig:CO-channels}.

\begin{figure*}
    \includegraphics[width=0.9\linewidth]{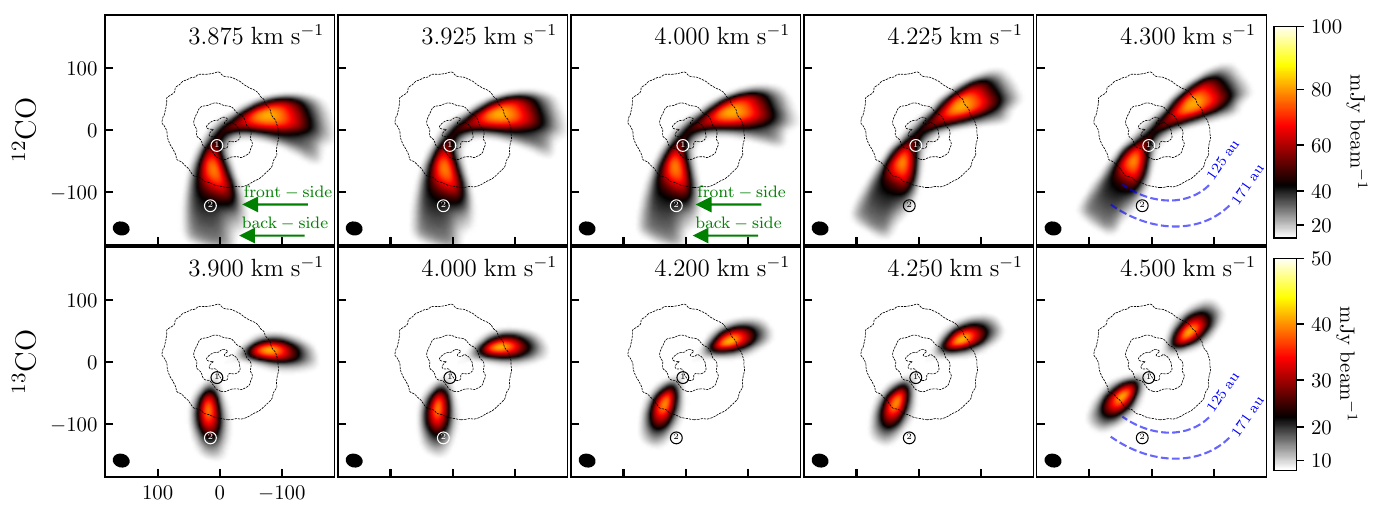}
    \includegraphics[width=0.9\linewidth]{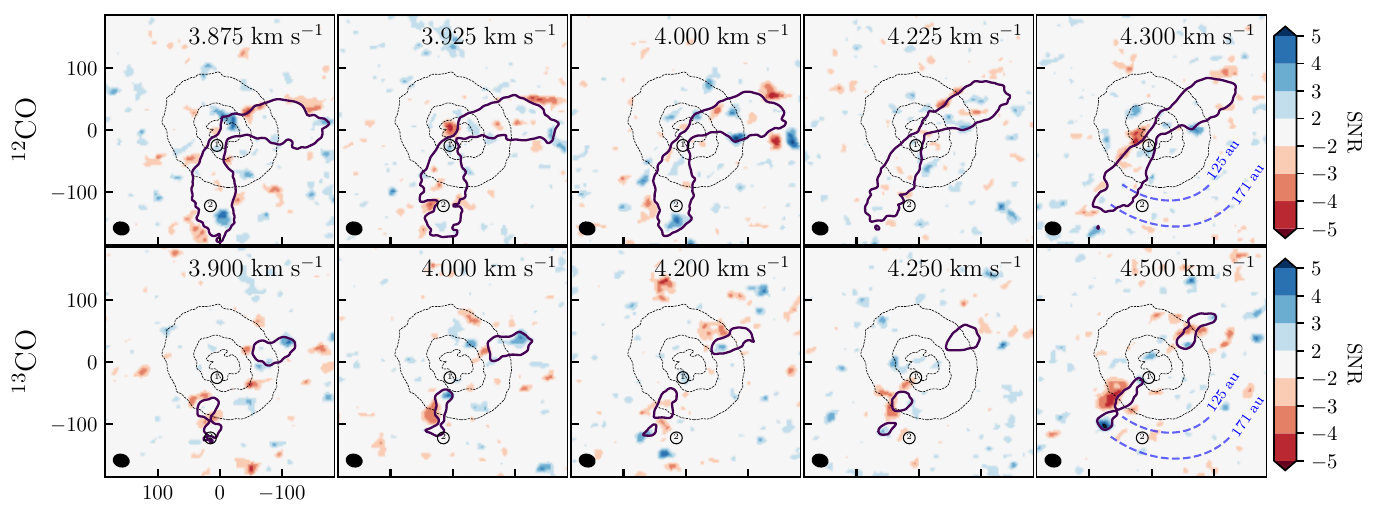}
    \caption{Model (top rows) and residual maps (bottom rows) after subtracting the best fit \textsc{discminer} model. The green arrows indicate the emission from the frontside and backside of the disc in a few channel maps, which may mimic kink-like features. The position of the inner planet candidate at $r = 32$ au is marked by \protect\textcircled{\textbf{1}}, while that of the outer planet candidate at $r = 148$ au is marked by \protect\textcircled{\textbf{2}}. The iso-contours show the ALMA Band 7 dust continuum emission at 7-$\sigma$ level.}
    \label{fig:CO-channels_Model}
\end{figure*}

\begin{figure*}
    \centering
    \includegraphics[width=0.9\linewidth]{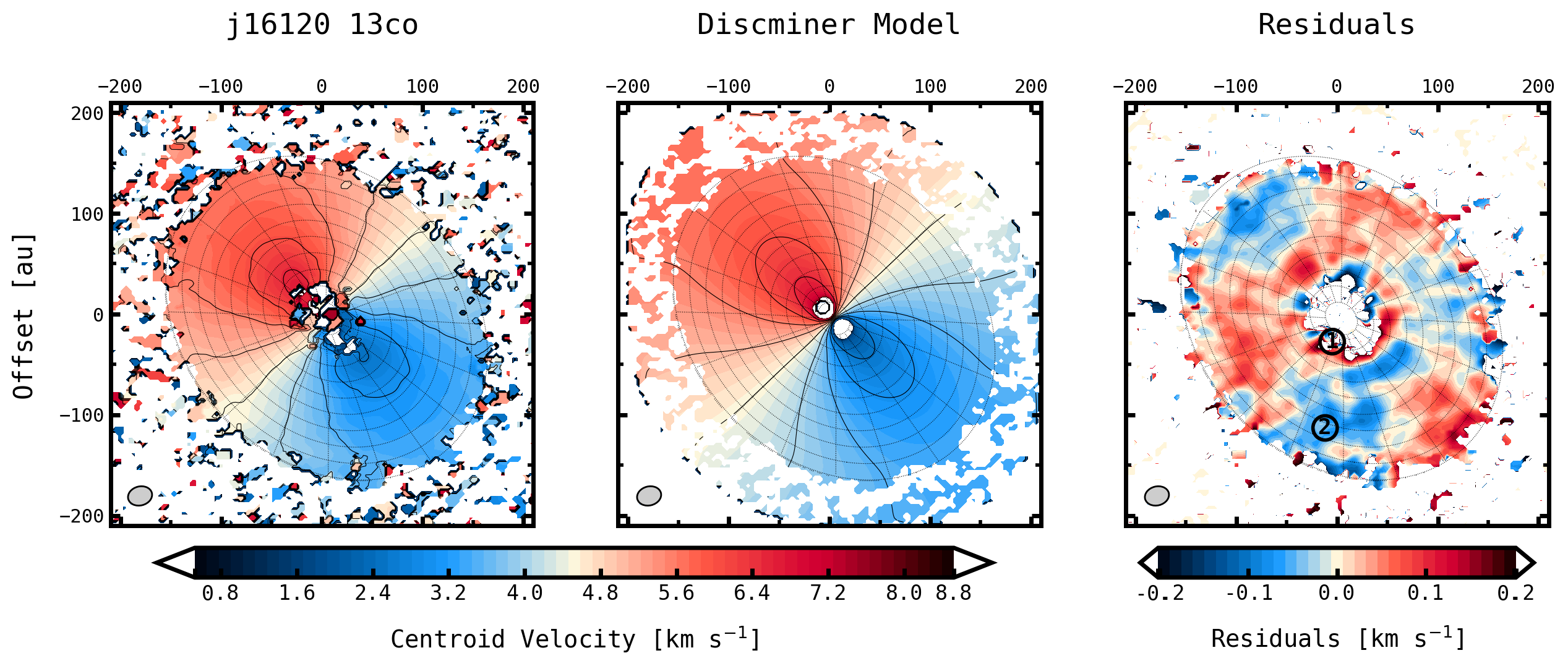}
    \caption{Same as Figure \ref{fig:CO-Discminer_Residuals} but for $^{13}$CO.}
    \label{fig:13CO-Discminer_Residuals}
\end{figure*}

The substructures observed between 125 and 171 au have corresponding residuals within $\pm 5\sigma$, making them compatible with the Keplerian \textsc{discminer} model.

Figure~\ref{fig:13CO-Discminer_Residuals} shows the kinematic model and residuals for $^{13}$CO. The sensitivity and angular resolution make it difficult to study the regions around the inner and outer planet candidates.

\section{Warped disc}\label{app:Wrap}

Figure~\ref{fig:Warp} shows the results of the velocity residual analysis using a warped disc model, obtained with the fitting scripts described in \cite{Winter_2025}. The model reproduces most of the observed red- and blue-shifted residuals, including the spiral-like structure associated with the inner planet candidate. The maximum difference in the disc inclination and position angle are $\sim 4, 3$ deg, respectively.

\begin{figure*}
    \centering
    \includegraphics[width=\linewidth]{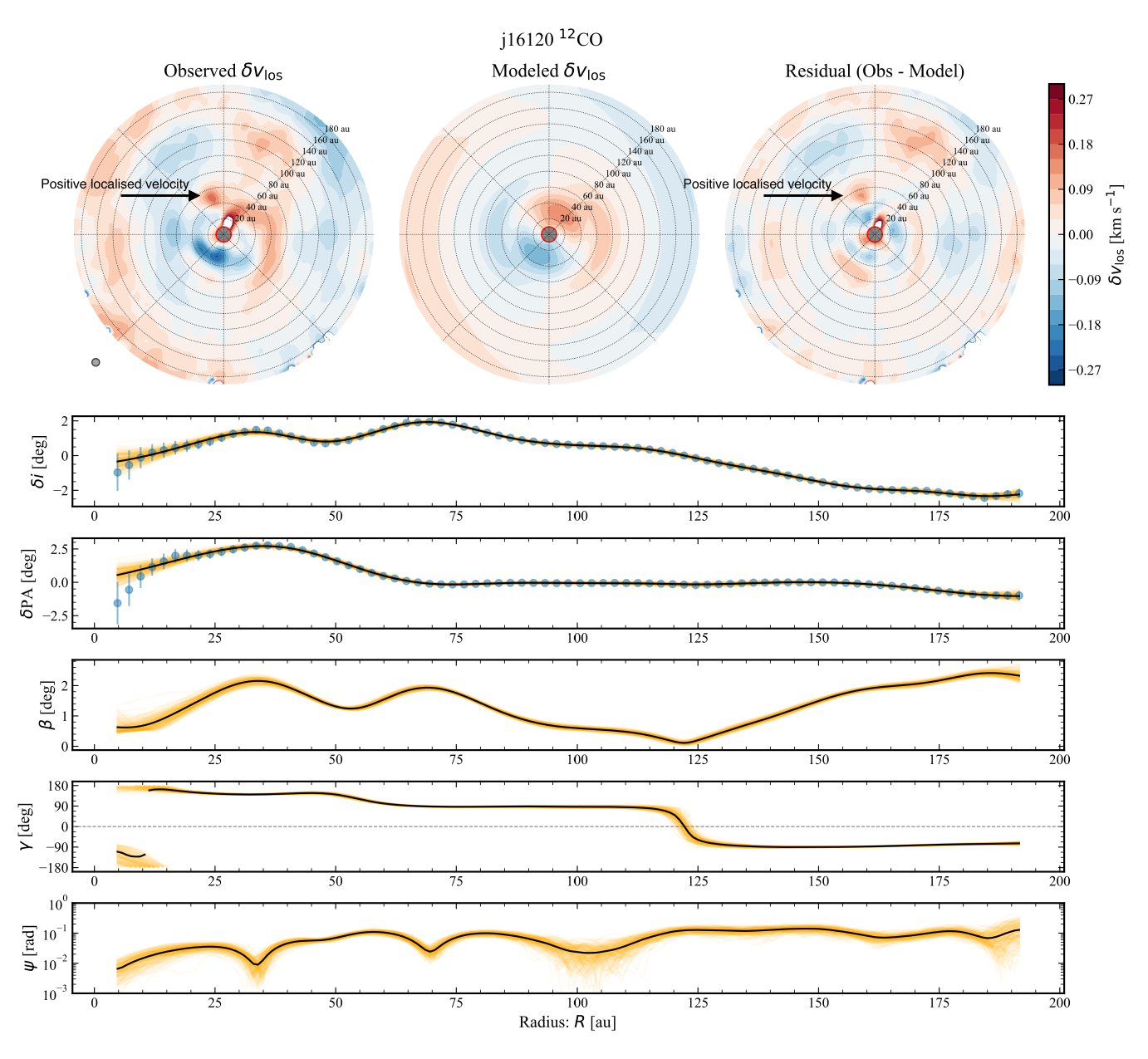}
    Top panels: Observed (left), Model (middle), and Residual (right) maps of the $^{12}$CO velocity residuals of J16120, assuming a warped disc. The arrow indicates the location of a positive localised velocity substructure that is not reproduced by the model.
    Radial profiles: Changes in the disc inclination ($\delta$i) and position angle ($\delta$PA), and the physical warp properties tilt $\beta$, twist $\gamma$, and the warp amplitude $\psi$.
    \label{fig:Warp}
\end{figure*}


\bsp	
\label{lastpage}
\end{document}